%% file: main.tex
\documentclass[twocolumn,
               trackchanges]{aastex63}
\usepackage[utf8]{inputenc}

\usepackage[T5,T1]{fontenc}
\usepackage{savesym}
\usepackage{amsmath}
\usepackage{booktabs}

\newcommand{\atel}[1]{\href{http://www.astronomerstelegram.org/?read=#1}{ATel {#1}}}

\shorttitle{IceCube Novae Search}

\begin{document}

\title{Search for sub-TeV Neutrino Emission from Novae with IceCube-DeepCore}

\input{authorlist120122.tex}

\noaffiliation

\begin{abstract}
The understanding of novae, the thermonuclear eruptions on the surfaces of white dwarf stars in binaries, has recently undergone a major paradigm shift. Though the bolometric luminosity of novae was long thought to arise directly from photons supplied by the thermonuclear runaway, recent GeV gamma-ray observations have supported the notion that a significant portion of the luminosity could come from radiative shocks. More recently, observations of novae have lent evidence that these shocks are acceleration sites for hadrons for at least some types of novae. In this scenario, a flux of neutrinos may accompany the observed gamma rays. As the gamma rays from most novae have only been observed up to a few GeV, novae have previously not been considered as targets for neutrino telescopes, which are most sensitive at and above TeV energies. Here, we present the first search for neutrinos from novae with energies between a few GeV and 10 TeV using IceCube-DeepCore, a densely instrumented region of the IceCube Neutrino Observatory with a reduced energy threshold. We search both for a correlation between gamma-ray and neutrino emission as well as between optical and neutrino emission from novae. 
We find no evidence for neutrino emission from the novae considered in this analysis and set upper limits for all gamma-ray detected novae.
\end{abstract}


\section{Introduction}
\label{sec:intro}
Novae are luminous outbursts that occur when a white dwarf in a binary system rapidly accretes matter from its companion star, leading to unstable nuclear burning on the surface of the white dwarf \citep{Gallagher:1978xe}. Surprisingly, it was recently discovered that these luminous events, typically identified in optical wavelengths, were often accompanied by gigaelectronvolt (GeV) gamma rays \citep{Ackermann:2014vfa}. Now, approximately one decade since the first discovery of gamma rays from novae with the Large Area Telescope (LAT) aboard NASA's \textit{Fermi} satellite, there have been over a dozen novae identified in gamma rays, some identified in archival searches \citep{Franckowiak:2017iwj}, and more recently being announced in real time via channels such as the Astronomer's Telegram (ATel)\footnote{\url{https://www.astronomerstelegram.org/}}. For a recent review of novae, see \cite{Chomiuk:2020zek}.

The gamma-ray emission from novae was recently found to be strongly correlated in time with the optical emission, lending evidence for a common origin in shocks \citep{Aydi:2020znu}. Novae have apparent efficiencies of nonthermal particle acceleration on the order of $0.3-1\%$~\citep{Li:2017crr,Aydi:2020znu}. Although low compared to efficiencies inferred for adiabatic shocks in Galactic accelerators like supernova remnants~\citep{Morlino:2011di}, novae still provide us with a relatively nearby class of transients that can be used to understand more energetic and distant classes of nonrelativistic, shock-powered transients~\citep{Fang:2020bkm}. The nonthermal gamma rays could, in principle, be produced from a variety of mechanisms, depending on the composition of the relativistic particles accelerated at the shocks. In the ``leptonic'' scenario, relativistic electrons Compton up-scatter the optical light or radiate via bremsstrahlung. In the ``hadronic'' scenario, accelerated relativistic ions interact with ambient gas, producing charged and neutral pions, which in turn decay into neutrinos and gamma rays, respectively. 

For many recent observed novae, several studies favor the hadronic scenario. For example, large magnetic fields are required to accelerate particles to produce the GeV gamma rays detected from some novae, which is inconsistent with leptonic models~\citep{Li:2017crr, Vurm:2016zea}. Moreover, recent observations of the spectrum of nova RS Ophiuchi during its 2021 outburst saw its spectrum extend beyond TeV energies, with fits to both H.E.S.S. and MAGIC data consistent with hadronic modeling~\citep{MAGIC:2022rmr,HESS:2022qap}.

In addition to modeling based on gamma-ray observations, neutrinos could hold the key to distinguishing between the leptonic and hadronic models~\citep{Razzaque:2010kp,Metzger:2015zka}. Furthermore, neutrinos could not only validate the hadronic scenario, but the level of any such neutrino flux would provide valuable information for understanding the environments of these shocks. 

However, while there is a guaranteed flux of neutrinos that would accompany any gamma rays produced in the hadronic scenario, the energies of the gamma rays place a limit on the maximum energies of neutrino counterparts. On average, hadronic gamma rays carry approximately 10\% of the initial relativistic ion energy, whereas the neutrinos carry about 5\% of the initial energy. Thus far, with the exception of nova RS Ophiuchi, the gamma-ray spectra from novae detected by the \textit{Fermi}-LAT have been modeled as a power law with an exponential cutoff, $dN/dE \propto E^{-\Gamma}\exp(-E/E_{\mathrm{cut}})$, where $\Gamma$ is the photon spectral index and $E_{\mathrm{cut}}$ is the cutoff energy \citep{Ackermann:2014vfa,Franckowiak:2017iwj}. Most individual novae have been fit with a cutoff of a few GeV, and a population analysis finds that the global spectrum from novae is well described with $\Gamma = 1.71 \pm 0.08$ and $E_{\mathrm{cut}} = 3.2 \pm 0.6$~GeV~\citep{Franckowiak:2017iwj}. While some novae show no strong evidence for a cutoff with LAT data alone, observations with MAGIC of an individual nova, V339 Del, place a strong upper limit in the TeV band, necessitating a cutoff in the 100 GeV range, if not at lower energies \citep{MAGIC:2015dda}. 

It is for this reason that novae have not previously been targeted by neutrino telescopes.~Optical Cherenkov neutrino telescopes, such as the IceCube Neutrino Observatory, typically have optimal sensitivities $\gtrsim$ TeV, which is significantly higher than one could expect from even the most optimistic models for cosmic-ray acceleration in novae.~Some previous searches for sub-TeV neutrino sources have been performed by Super-Kamiokande~\citep{Super-Kamiokande:2009uwx,Super-Kamiokande:2018dbf} as well as by IceCube~\citep{IceCube:2015wtd,IceCube:2020qls}, but these analyses have targeted other classes of astrophysical transients. 

In this work, we describe the first search for neutrinos from novae using IceCube-DeepCore, a subarray of the IceCube Neutrino Observatory, which can be used to lower the energy threshold for astrophysical neutrino source searches. Although DeepCore has been used in past all-sky searches for generic transients \citep{IceCube:2015wtd,IceCube:2020qls}, this is the first time it has been used in a search for Galactic transients.

We begin in section~\ref{sec:novae_sample} by describing the sample of novae used in this work, distinguishing those detected by the \textit{Fermi}-LAT from those only detected in optical wavelengths. In section~\ref{sec:greco_icecube}, we describe DeepCore in more depth, and list the details of the event selection used in our analyses. Then, in section~\ref{sec:analysis}, we describe the maximum likelihood approaches we implemented, including, in section~\ref{sec:gamma_search}, a search for correlations with individual gamma-ray light curves, and in section~\ref{sec:optical_stacking}, a search for a cumulative signal from all optically detected novae. We summarize our results in section~\ref{sec:results}, and then discuss the implications of our results as well as of using DeepCore for neutrino astronomy in section~\ref{sec:discussion}.

\section{Nova sample}
\label{sec:novae_sample}
To create our list of novae, we began by using the catalog compiled for a search for gamma-ray emission from novae in \cite{Franckowiak:2017iwj}. This catalog spanned the years 2008\textendash 2015, whereas the neutrino data used in our analysis does not begin until 26 April 2012 and extends until 29 May 2020. To gather information on more recent novae as well as to ensure that the catalog of all novae up to 2015 was exhaustive, we cross-referenced our sample against other compilations.\footnote{\url{https://asd.gsfc.nasa.gov/Koji.Mukai/novae/novae} by Koji Mukai}$^{,}$\footnote{\textit{List of Galactic Novae}: \url{http://projectpluto.com/galnovae/galnovae.txt}, by Bill Gray} For additional information on gamma-ray novae, we searched ATels, and include two more recent novae detected by the LAT at the >5$\sigma$ level, V3890 Sgr and V1707 Sco. All gamma-ray detected novae used in this search are detected by the LAT at the 3$\sigma$ level or above, with the exceptions of V745 Sco and V1535 Sco, which were only identified as ``candidate'' gamma-ray sources in \cite{Franckowiak:2017iwj}, because they did not reach a detection at the 3$\sigma$ level, but are included here for completeness. 
For the gamma ray detected novae, we use the same time window for the neutrino search as the LAT detection window, as specified in Table \ref{tab:gamma_results}.
Time windows for the majority of the novae match those used in \cite{Gordon:2020fqv}. For the ``candidate'' novae V745 Sco and V1535 Sco, we use the times reported in \cite{Franckowiak:2017iwj}. For the more recent novae V3890 Sgr and V1707 Sco, we use the detection times reported in their respective ATels\footnote{\atel{13114}}$^,$\footnote{\atel{13116}}.

\begin{figure}
    \centering
    \includegraphics[width=0.45\textwidth]{ 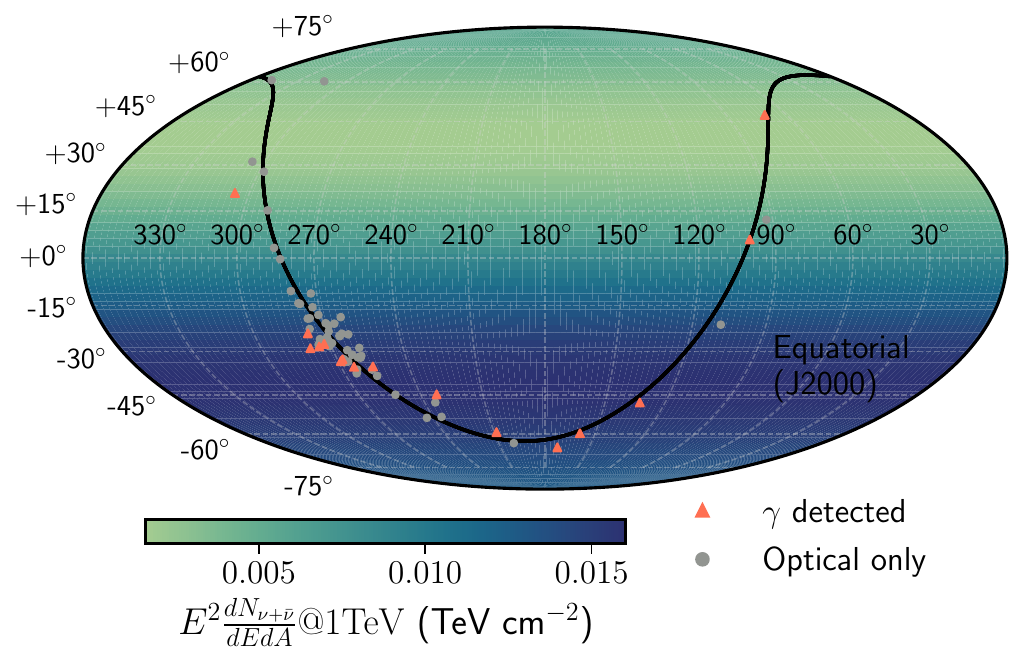}
    \caption{Locations of novae investigated in this work. Those novae that were identified in GeV gamma rays are shown in orange, while those only detected at other wavelengths are shown in gray. These novae are shown on top of the sensitivity (described fully in Section~\ref{sec:analysis}) our unstacked analysis (described in Section~\ref{sec:gamma_search}) would have to a single nova, assuming an analysis time window of 1 day.}
    \label{fig:skymap_with_sens}
\end{figure}

Additionally, for all novae we require a measurement of the optical peak time. For novae through 2015, this was explicitly calculated in \cite{Franckowiak:2017iwj}. For later novae, we accessed light curve information through AAVSO\footnote{\url{https://www.aavso.org/LCGv2/}} as well as through the Stony Brook/SMARTS Spectroscopic Atlas of Southern Novae\footnote{\url{http://www.astro.sunysb.edu/fwalter/SMARTS/NovaAtlas/atlas.html}}, and, where available, we also included observations from the ASAS-SN transient server \citep{Shappee:2013mna}. If a nova was the subject of a dedicated publication that included optical light curves, we took the information from the respective work. As some novae are believed to be discovered after peak brightness, and as some novae have light curves with only sparse sampling, we are not able to precisely determine the date of optical peak brightness; 
therefore, the novae V1404 Cen, V5854 Sgr, V1657 Sco, V3663 Oph, FM Cir, V3731 Oph, V3730 Oph, V2891 Cyg, and V1709 Sco are excluded from our analysis.

Our final sample contains 67 novae, 16 of which were detected by the LAT at GeV energies. The full table of all of the novae used in this work, including the dates of peak optical brightness as well as periods of gamma-ray detection, is included in Table \ref{tab:novae_cat} of Appendix~\ref{sec:catalog_appendix}. Figure~\ref{fig:skymap_with_sens} displays the locations of these novae superimposed upon the sensitivity of the analysis, which will be described in Section~\ref{sec:gamma_search}. As expected, the novae are most commonly found near the Galactic plane, with a large number in the Galactic bulge region. It is worth noting that nova RS Ophiuchi is not included in this sample. As there is only low-energy neutrino data processed through 2020, there is no synchronous data available for an analysis of the emission seen from nova RS Ophiuchi in its 2021 outburst; however, this nova can be included in future analyses with more years of processed data.

\section{IceCube-DeepCore}
\label{sec:greco_icecube}
The IceCube Neutrino Observatory is a gigaton-scale ice Cherenkov detector at the geographic South Pole~\citep{IceCube:2016zyt}. The detector consists of 86 vertical strings, each of which is instrumented with 60 digital optical modules (DOMs) that house 10-inch photomultiplier tubes. Neutrinos are detected indirectly via the Cherenkov radiation produced from relativistic charged particles created by neutrino interactions in the surrounding ice or nearby bedrock beneath IceCube. For the majority of the strings in the detector, the DOMs are evenly spaced between 1.45~km to 2.45~km below the surface of the ice, and the strings are arranged in a hexagonal grid. 
In the main array, the strings have a horizontal spacing of 125~m and the DOMs are spaced 17~m apart on each string, which optimizes the sensitivity of the detector at energies in the $\sim$TeV$-$PeV range, and has some sensitivity down to $\sim$100~GeV. In the center of the array, there is a more densely instrumented region of 8 specialized low-energy strings, with reduced spacings of both the strings (72~m apart) and the DOMs on each string (7~m apart), and the photomultiplier tubes in these DOMs have higher quantum efficiency than those in the main array. These 8 strings, as well as the 7 innermost strings of the main array, make up IceCube-DeepCore~\citep{IceCube:2011ucd}, and its reduced spacing and higher quantum efficiencies allow a reduced energy threshold down to $\sim$10~GeV.

In this analysis, we use a modified version of the GRECO (GeV Reconstructed Events with Containment for Oscillation) dataset that was developed for oscillation studies with DeepCore, as described in~\cite{IceCube:2019dqi}, which we denote as the GRECO Astronomy dataset for the remainder of this paper. This event selection is described in detail in Appendix~\ref{app:greco}, including the selection criteria, the event selection response as a function of energy, as well as a description of the angular uncertainty of the events that are used in the sample. In short, the effective area is similar across the whole sky, and the selection is sensitive to neutrino emission in the energy range from a few GeV to about 10 TeV, depending on the assumed spectral hypothesis of a source. Depending on the energy and type of event morphology, events in the GRECO Astronomy dataset can have uncertainties in the angular resolution ranging from a few degrees to approximately 80$^{\circ}$, as shown in Figure~\ref{fig:greco_err} (Appendix~\ref{app:greco}). The events used in this selection have angular uncertainties that are much larger than those present in $\gtrsim$ TeV neutrino analyses, because the track lengths of outgoing muons (if present) are shorter in DeepCore, and shorter track lengths increase angular uncertainty.

\section{Analysis Techniques}
\label{sec:analysis}
The analyses performed for this work rely on an unbinned maximum likelihood approach that is common in searches for astrophysical neutrino sources~\citep{Braun:2008bg}. For this work, we implement an ``extended likelihood'' approach, which splits the entire dataset into a time period of interest (the ``on-time window'') with duration $\Delta T$ and the remainder of the dataset (the ``off-time window''), with $\Delta T$ defined by the duration of the electromagnetic outburst for each nova. This approach has been used in many other searches for short-timescale neutrino transients~\citep{2017ApJ...843..112A, IceCube:2019acm, 2020ApJ...898L..10A}. 

For an on-time window with $N$ total neutrino candidate events, we define the likelihood, $\mathcal{L}$, which consists of the sum of a signal probability density function (PDF), $\mathcal{S}$, and a background PDF, $\mathcal{B}$, as

\begin{equation}
\begin{aligned}
    \label{eq:likelihood}
    \mathcal{L}(n_{s}, &\gamma) = \frac{\left(n_{s}+n_{b}\right)^{N} e^{-\left(n_{s}+n_{b}\right)}}{N !} \\ & \times \prod_{i=1}^{N}\left[ \frac{n_{s}}{n_{s}+n_{b}} \mathcal{S}\left(x_{i} | \gamma \right)+\frac{n_{b}}{n_{s}+n_{b}} \mathcal{B}\left(x_{i}\right)\right],
\end{aligned}
\end{equation}
where $n_s$ and $n_b$ are the signal and expected background event counts, respectively, and $\gamma$ is the spectral index of the source, as defined in Equation \ref{eq:dnde} below. The index $i$ iterates over all neutrino candidate events in the on-time window, and each of these events has a set of observables $x_i$. The term outside of the product in the likelihood accounts for possible fluctuations in the rate of events, which is helpful in transient-style analyses where the expected number of background events is small \citep{1990NIMPA.297..496B}. In order to calculate the expected number of events, we assume a neutrino number flux with a power-law spectrum: 
\begin{equation}
    \label{eq:dnde}
    F_{\nu+\bar{\nu}}(E | \gamma) = \frac{\mathrm{d}N_{\nu+\bar{\nu}}}{\mathrm{d}E\,\mathrm{d}A\,\mathrm{d}t} \Delta T = \phi_0 \left( \frac{E}{E_0} \right)^{-\gamma}
\end{equation}
where $\gamma$ is the spectral index, $\phi_0$ the flux normalization, at a reference energy $E_0=\mathrm{1\,TeV}$. We maximize the likelihood with respect to both $n_s$ and $\gamma$, to characterize the normalization and spectrum of neutrino emission from a potential source.

Both the signal and background PDFs consist of terms describing the energy and the spatial distributions. Beginning with the signal PDF, $\mathcal{S}$, the spatial component expects neutrinos from novae to be spatially associated with the nova location observed from photons. We define the PDF as it is used in~\cite{IceCube:2017amx}, which is a function of the opening angle between the source location and the reconstructed event direction, $\Delta \Psi$, and which can be approximated by the first-order non-elliptical component of the Kent distribution. The Kent distribution is an extension of a radially symmetric Gaussian that accounts for projection effects onto a sphere, which is necessary when handling events with large uncertainties in the reconstructed direction. The width of this distribution is governed by a per-event angular uncertainty, $\sigma_i$. The energy term of the signal PDF is a function of each event's best-fit decl. and energy, and is calculated for a variety of possible source spectral indices. It is described fully in~\cite{IceCube:2016tpw}.

For the background PDF, $\mathcal{B}$, both the energy and spatial components are parameterized directly from experimental data. The spatial component depends only on each reconstructed event's decl., and the energy component is a function of the event's reconstructed decl. and its reconstructed energy. The probability in R.A. is treated as a uniform distribution. The number of expected background events, $n_b$, is calculated from off-time data to obtain an estimate on the all-sky rate of the sample.

The final test statistic (TS) is based on a log likelihood ratio between the best-fit signal hypothesis (with best-fit parameters $\hat{n}_s, \hat{\gamma}$) and the background-only hypothesis ($n_s=0$, for which $\gamma$ is undefined), and simplifies to

\begin{equation}
\label{eq:TS}
    \mathcal{TS}=-2\hat{n}_{s}+2\sum_{i=1}^{N} \ln \left[\frac{\hat{n}_{s} \mathcal{S}\left(x_{i}| \hat{\gamma} \right)}{n_{b} \mathcal{B}\left(x_{i}\right)}+1\right] \; .
\end{equation}

\subsection{Gamma-ray correlation analysis}
\label{sec:gamma_search}
The first analysis of this work looks for neutrino emission coincident with gamma-ray emission. For each nova that has been detected in gamma rays (or was identified as a gamma-ray candidate in~\cite{Franckowiak:2017iwj}), we use the time period of gamma-ray observations to determine the on-time window for the analysis. The time windows are consistent with what was reported for gamma-ray activity in~\cite{Gordon:2020fqv} and with the time windows for the candidate emitters in~\cite{Franckowiak:2017iwj}. The time windows range from days to several weeks in duration. As the GRECO Astronomy dataset can only be used to search for transient neutrino sources due to high atmospheric backgrounds at these energies, we validated that we could search for neutrino emission on all of these timescales by injecting artificial signal events into our analysis and making sure that the analysis could recover an injected signal. One way of characterizing the analysis performance is by calculating the sensitivity, which is defined as the median one-sided Neyman upper limit (at the 90\% confidence level) that one could expect to set under the background-only hypothesis. In Figure~\ref{fig:skymap_with_sens}, we show the analysis sensitivity at all locations on the sky for an on-time window of 1 day, which is the minimum time window used in the gamma-ray correlation analysis. The sensitivity is better in the Northern Sky, where the atmospheric backgrounds are smaller, but the variation in analysis performance between the Northern and the Southern Sky is not as extreme as in higher-energy analyses, where the analysis performance can differ by orders of magnitude in the two celestial hemispheres. 

Then, for each of the 16 gamma-ray novae with gamma-ray observations that were coincident with the GRECO Astronomy dataset, we perform the likelihood analysis as described above. In addition to calculating the best-fit parameters and test statistics, we also calculate $p$-values for quantifying the compatibility with the background-only hypothesis. These $p$-values are one-sided and are calculated by comparing the observed test statistic to an expected distribution of test statistics created by performing pseudo-experiments assuming the background-only hypothesis.

\begin{figure}
    \centering
    \includegraphics[width=0.47\textwidth]{ 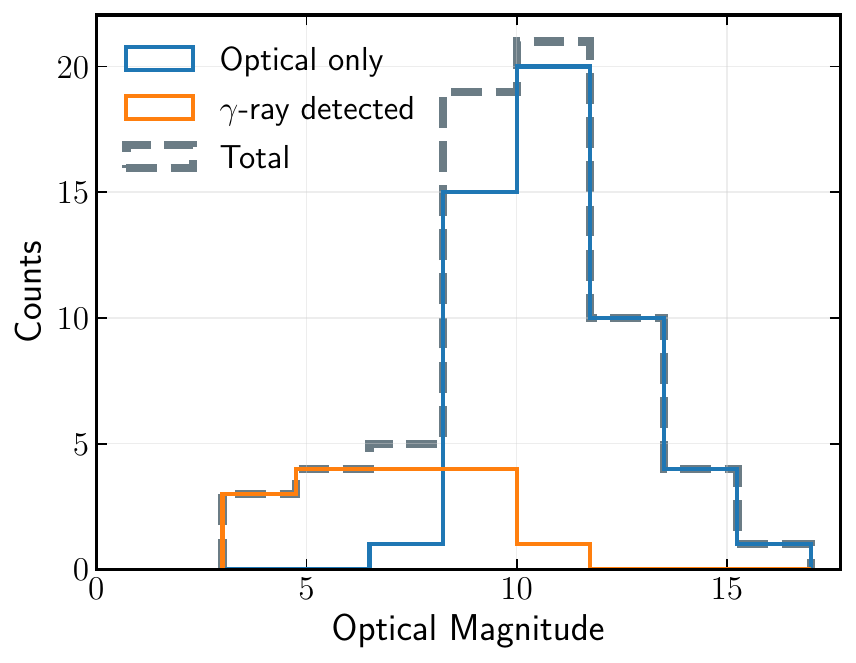}
    \caption{Peak optical magnitudes of the novae used in this analysis. Those novae that were detected in gamma rays or were identified as candidate gamma-ray emitters in~\cite{Franckowiak:2017iwj} are shown in the orange histogram, whereas the blue histogram shows novae that were not detected in gamma rays. The gray histogram shows the total of all novae used in the analysis.}
    \label{fig:optical_mag}
\end{figure}

\subsection{Stacking analysis}
\label{sec:optical_stacking}
The second analysis of this work stacks together the contribution from multiple novae in order to improve the sensitivity of the search. For this construction, the PDFs in the likelihood are weighted sums of the PDFs from each individual source, as in~\cite{IceCube:2017amx}. While stacking analyses have the advantage of improving the average sensitivity of the sample, they require testing a particular hypothesis about the distribution of fluxes from the individual sources. There is no conclusive or agreed upon model for calibrating neutrino fluxes based on observations of novae across the electromagnetic spectrum. 
However, as the production of neutrinos is intimately linked to the production of gamma rays, the gamma rays could provide a handle on the expectation for a corresponding neutrino flux. If, on the other hand, the shocks that are producing neutrinos are optically thick for gamma-ray emission, then any potential neutrino flux could overshoot the corresponding gamma-ray flux~\citep{Bednarek:2022vey}. 
In this case, the optical emission could provide the best handle on the level of a neutrino flux, as it has been suggested that the total energy that goes into particle acceleration could be correlated with the optical energy output~\citep{Fang:2020bkm}. 

We therefore test two different weighting schemes: (1) gamma-ray stacking and (2) optical stacking. For gamma-ray stacking, we use the 16 novae that were detected in gamma rays, and weight them based on the normalization of the gamma-ray flux at 100~MeV that was fit by \textit{Fermi}-LAT. For the optical stacking weighting scheme, we no longer require that a nova be detected in gamma rays. Figure~\ref{fig:optical_mag} shows the distribution of peak optical magnitudes for all novae in the sample. We weight the novae based on their peak optical fluxes, which are converted from the optical magnitudes. We break this distribution up into novae that were detected in gamma rays and those that were not. In general, the novae that were detected in gamma rays were brighter optically, which might be an indication that all novae are gamma-ray emitters, but some are at a level that is too dim to be detected by \textit{Fermi}-LAT.

\begin{figure*}[t]
    \centering
    \includegraphics[width=0.95\textwidth]{ 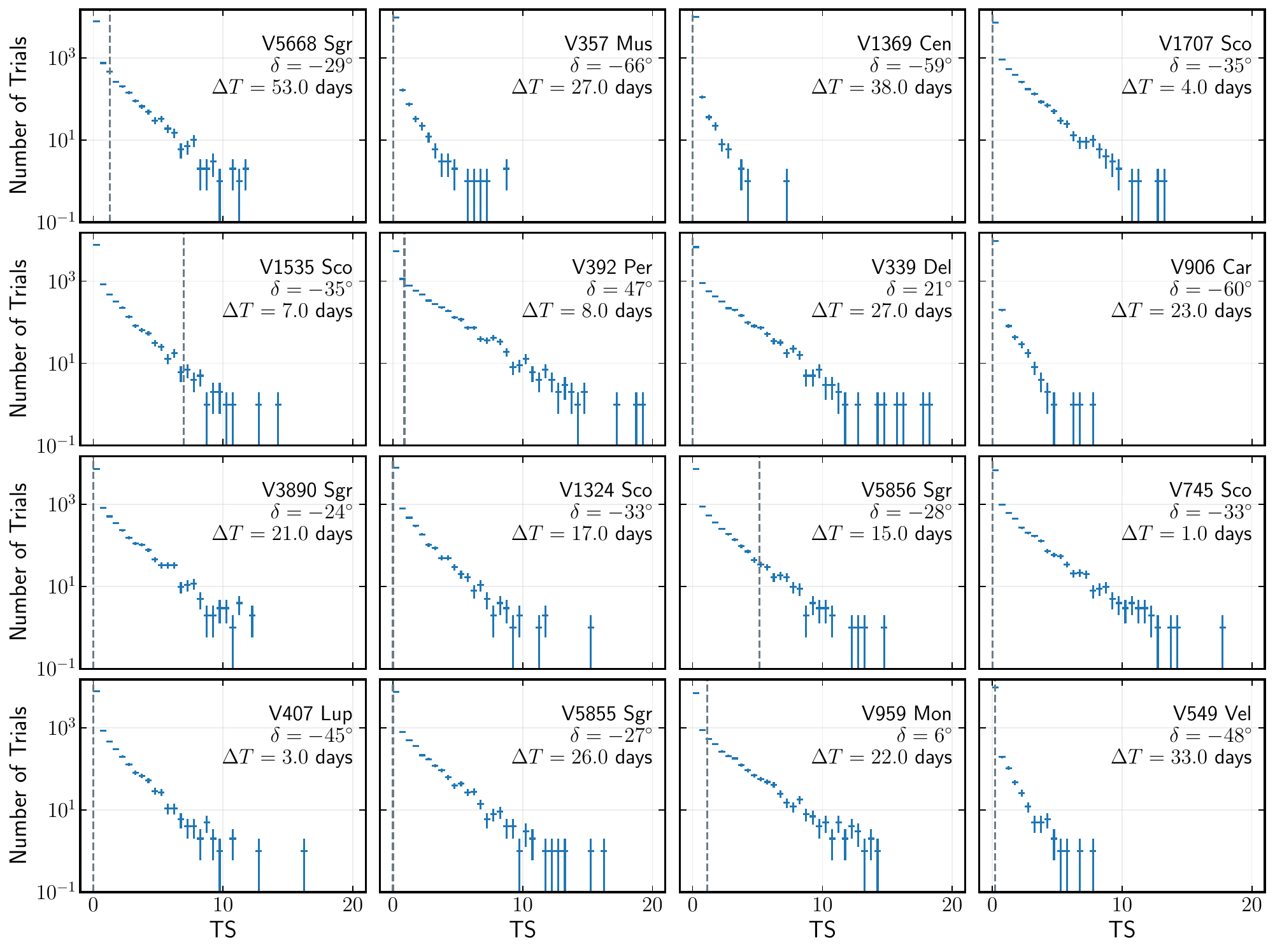}
    \caption{Comparison of observed test-statistics (gray dashed lines) to expectations from 10,000 pseudo-experiments performed under the assumption of the null hypothesis (blue). Each panel represents the gamma-ray correlation analysis for an individual nova.}
    \label{fig:ts_dists}
\end{figure*}

\input{gamma_ray_results}

We also require an on-time window for each of the novae in both of these weighting schemes. As mentioned above, using the GRECO Astronomy dataset to search for point sources is only feasible when searching for transient emission. When stacking together sources, this becomes especially important because, if there are $k$ sources each with an on-time window of $\Delta T$, the effective background is equivalent to a single source with a time window of $k \Delta T$. We find that when stacking sources, this limits us to searching for emission within a smaller time window than we could use for the gamma-ray correlation analysis. After injecting artificial signal into the analysis over a variety of time windows, we find that the analysis is able to precisely recover the amount of injected signal for on-time windows of 1 day or less. To maximize the amount of signal that we can integrate over, both weighting schemes used $\Delta T = 1~\mathrm{day}$ for the duration of the stacking analyses. For many of the gamma-ray novae, the gamma-ray light curves are not resolved enough to isolate the peak day of emission. Additionally, light curves of recently detected bright gamma-ray novae, such as V906 Car and ASASSN-16ma, show correlations between the optical and gamma-ray light curves, with delays between the two light curves of significantly less than 1 day. For these reasons, we center the time window on the date of peak optical emission for both the gamma-ray stacking and optical stacking analyses. These dates, as well as the peak optical magnitudes, are included in Table~\ref{tab:novae_cat} in Appendix~\ref{sec:catalog_appendix}.

For the gamma-ray correlation analysis, we find that to be sensitive to an individual source with a spectrum $F_{\nu+\bar{\nu}}(E | \gamma =2.0)$ in the Northern Sky with an on-time window of 1 day, we require 5 neutrinos. On the other hand, an observation of a source with the same spectrum in the Southern Sky with a long on-time window would require around 50 neutrinos. When we stack sources together for the gamma-ray weighting scheme, we find that we would need around 16 (53) signal events for an $F_{\nu+\bar{\nu}}(E | \gamma =2.0)$ ($F_{\nu+\bar{\nu}}(E | \gamma =3.0)$) spectrum, whereas for the optical stacking weighting scheme we would need around 27 (130) signal events for an $F_{\nu+\bar{\nu}}(E | \gamma =2.0)$ ($F_{\nu+\bar{\nu}}(E | \gamma =3.0)$) spectrum. In terms of per-source contributions, this means, on average, reducing the requirement from multiple detected neutrinos from each nova to 1 (0.4) detected neutrinos from each nova for the gamma-ray (optical) weighting scheme.

\subsection{Systematic Uncertainties}
\label{sec:systematics}
We estimate the effects of systematic uncertainties on our results by varying the most important systematic effects in IceCube. These systematics include ice properties, such as scattering and absorption of photons in the ice; relative DOM efficiency; and properties of the refrozen column of ice around the DOMs resulting from the drilling process, known as ``hole ice.'' In order to calculate the effect of systematic uncertainties on our upper limits, we produced two additional Monte Carlo datasets for each systematic effect, with a discrete value of the systematic chosen to bracket the uncertainty of that systematic effect. We simulate systematics datasets with $\pm$10\% DOM efficiency, $\pm$10\% absorption coefficient, $\pm$10\% scattering coefficient, and $\pm$1 $\sigma$ variations in hole ice optical properties, which were chosen to match those used in \cite{IceCube:2020qls}.

\begin{figure*}
    \centering
    \includegraphics[width=0.95\textwidth]{ 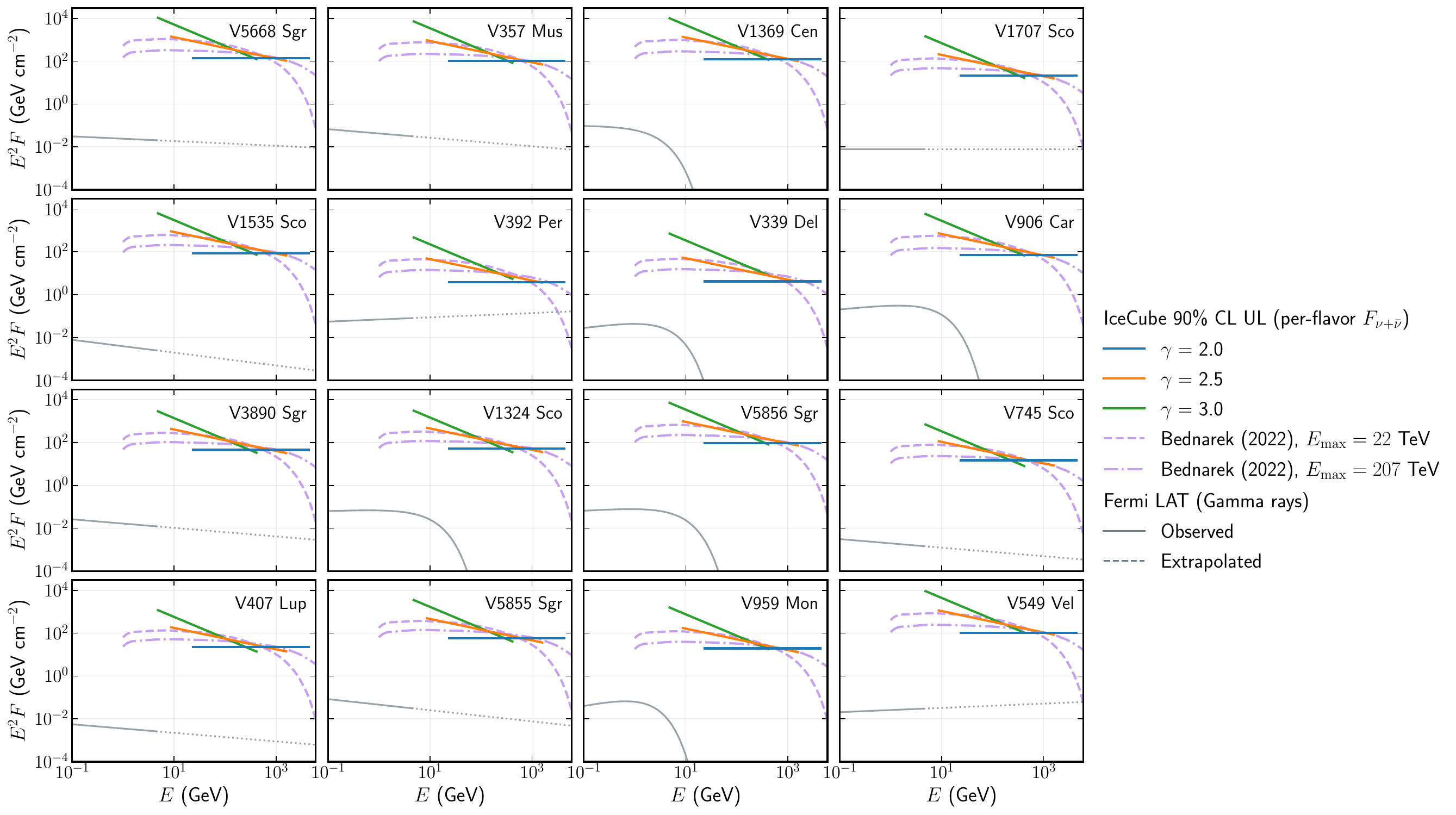}
    \caption{Upper limits on neutrino emission from all of the novae analyzed in the gamma-ray correlation analysis. Upper limits on power laws span the 90\% energy ranges discussed in Figure~\ref{fig:central_energies}, and include systematic effects as discussed in Section~\ref{sec:systematics}. In addition to constraining power laws, we also inject the spectral shapes from models in~\cite{Bednarek:2022vey}, and our upper limits on those spectra are shown in pink. We compare these fluxes to the measured gamma-ray fluxes (gray). For those gamma-ray detected novae that did not show evidence for a cutoff in the gamma-ray spectra, we show the lines as dotted past the global cutoff energy found in~\cite{Franckowiak:2017iwj}.}
    \label{fig:ul_panel}
\end{figure*}

We recalculate our sensitivities for each nova at 3 different spectral indices ($\gamma=$2.0, 2.5, and 3.0) using each of these datasets. We find the ratio of the sensitivity with the systematic included to the sensitivity without the systematic included, and average this value over all novae for each systematic. We then sum the components of these uncertainties in quadrature to find a total systematic uncertainty for each of the three spectral indices. We find the total systematic uncertainty on the flux upper limit, averaged over novae, to be within $\pm$21\% for a spectral index of $2.0$, $\pm$20\% for an index of $2.5$, and $\pm$13\% for an index of $3.0$. We rescale our power-law upper limits presented in Figure \ref{fig:ul_panel} using the upper bound of these calculated systematic uncertainties for each spectral index, to provide a conservative upper limit. 

\section{Results}
\label{sec:results}
After calculating test statistics for each analysis on true experimental data, no significant signal is detected, either in the gamma-ray correlation analysis or in either of the stacked analyses. 

In Table~\ref{tab:gamma_results}, we include all of the results from the gamma-ray correlation analysis. We also show the observed test statistics for each of the gamma-ray detected novae, as well as the background test statistic distribution calculated by performing pseudo-experiments assuming a background-only hypothesis in Figure~\ref{fig:ts_dists}. In addition to the time windows of the analysis, we include the best-fit parameters and pre-trial $p$-values, before accounting for the trials correction factor accrued from analyzing multiple sources. The most significant follow-up comes from our investigation of the nova V1535~Sco, which had a pre-trials $p$-value of $2.4\times 10^{-3}$, which gives a post-trials $p$-value of $3.8\%$ after accounting for the number of novae investigated, which we find to be consistent with expectations from atmospheric backgrounds. Nova V1535~Sco was not significantly detected in gamma rays, and was only identified as a candidate gamma-ray emitter in~\cite{Franckowiak:2017iwj}.

\subsection{Power-law Upper Limits}

As no significant signal was detected, we calculate upper limits on neutrino fluxes from novae during time periods of coincident gamma-ray detections. These limits are displayed in Figure~\ref{fig:ul_panel}. We also include a comparison to the gamma-ray fluxes as measured by \textit{Fermi}-LAT. When comparing the levels of the observed gamma-ray fluxes and the upper limits set by this analysis, it becomes apparent that more sensitive detectors might be required before one can expect to detect a neutrino signal from novae. 

\begin{figure*}
    \centering
    \includegraphics[width=0.46\textwidth]{ 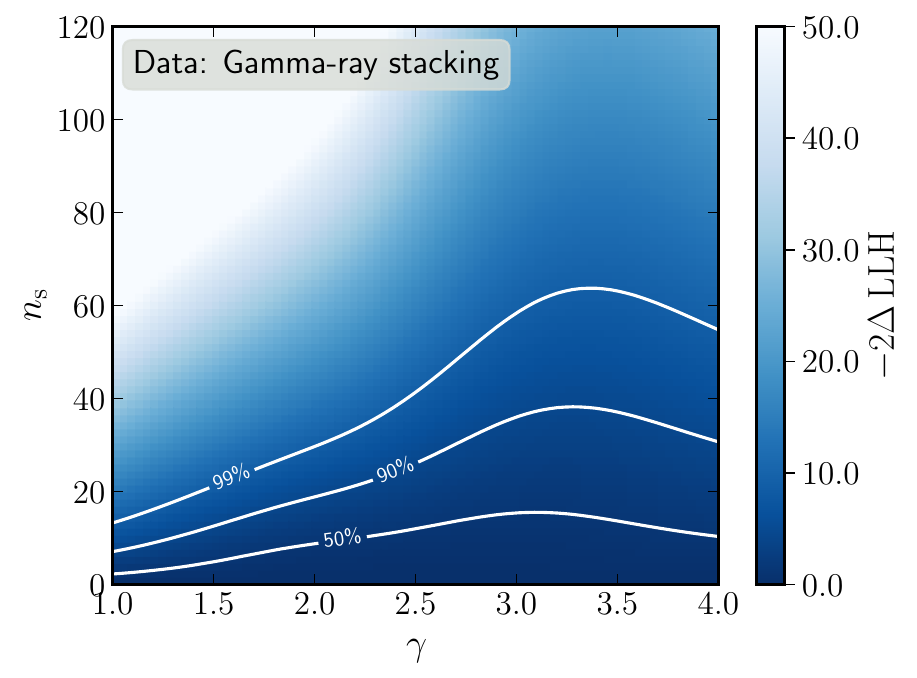}
    \includegraphics[width=0.46\textwidth]{ 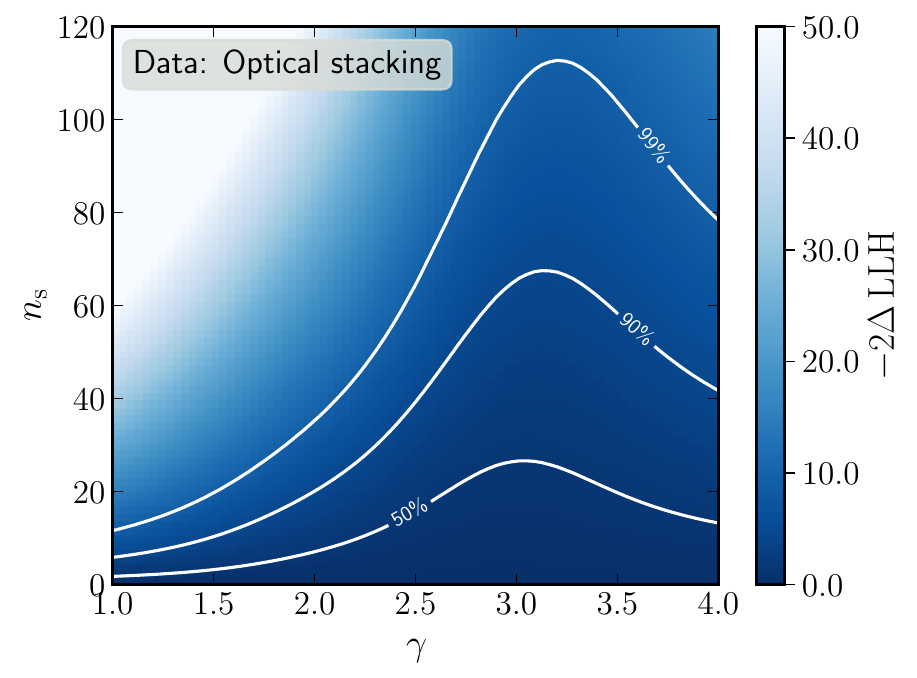}
    \includegraphics[width=0.46\textwidth]{ 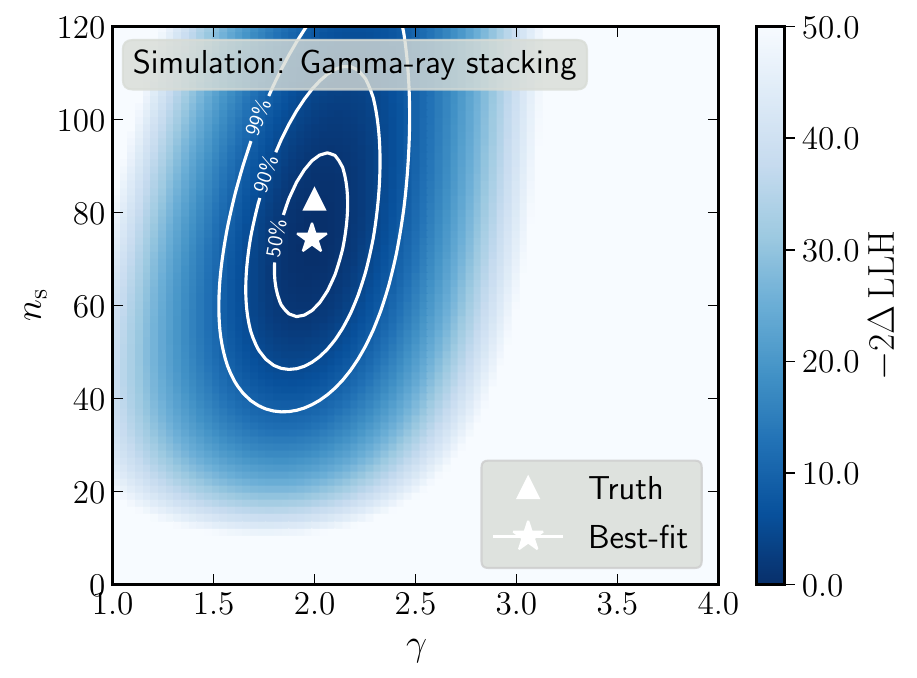}
    \includegraphics[width=0.46\textwidth]{ 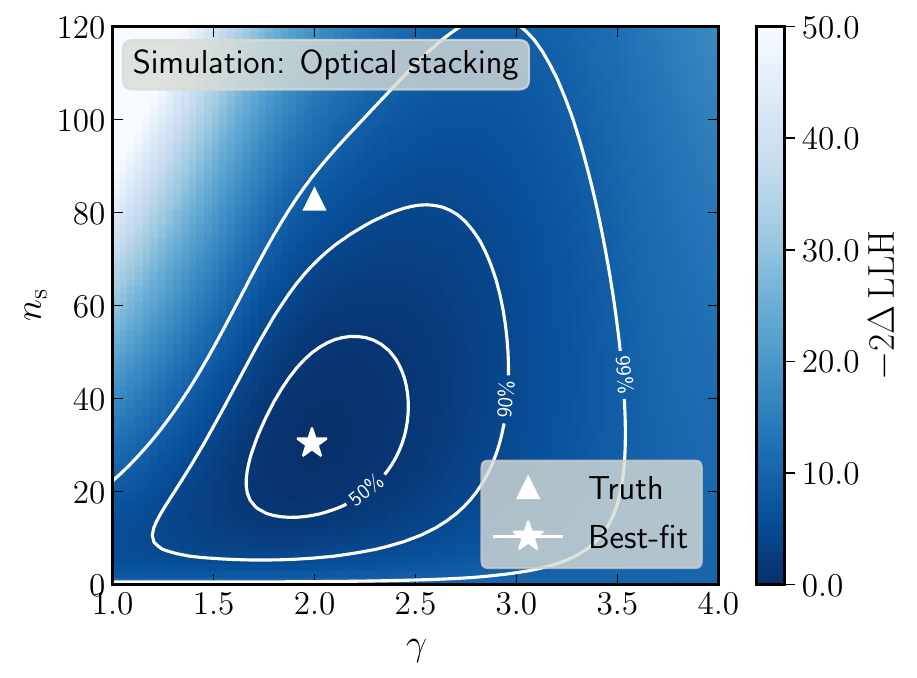}
    \caption{Stacking likelihood spaces for the gamma-ray stacking analysis (left) and for the optical stacking analysis (right). Here, $n_s$ refers to the total signal events for the entire stacked sample. The top row shows our observed data, and the bottom row shows a sample pseudo-experiment with injected signal. Contours denote 50\%, 90\%, and 99\% containment assuming Wilk's theorem with two degrees of freedom. For our observed data (top), $\hat{\gamma}$ is undefined when $\hat{n}_s = 0$, so we do not include a best-fit point on these panels.}
    \label{fig:llh_spaces}
\end{figure*}

Similar to the gamma-ray correlation analysis, no significant signal was detected in the stacking analysis, for either the gamma-ray weighting scheme or the optical stacking scheme. The likelihood maximization for each of the weighting schemes finds a best-fit of $\hat{n}_s = 0$, which yields $\mathcal{TS}=0$, and a pre-trial $p$-value of 1.0 for each of these weighting schemes. In Figure~\ref{fig:llh_spaces}, we show the likelihood spaces for our observed data, and we compare this against simulations in which we injected artificial signal, to contrast how our observed data appears relative to what we might expect for a bright signal from novae. 
In the simulation panels, we show how well the properties of the injected signal can be recovered by the analysis. There is a tendency for the analysis to underestimate the signal, which is especially visible in the optical stacking (Figure~\ref{fig:llh_spaces}, lower-right panel). This predominantly comes from the large angular uncertainties of events in the GRECO Astronomy dataset (see Appendix~\ref{app:greco}, Figure~\ref{fig:greco_err}), which causes low-energy signal events to appear background-like, and the assumption that the uncertainty contours are symmetric and Gaussian. This bias does not affect any constraints we set, as we calculate our upper limits using the one-sided Neyman construction, which handles this bias self-consistently.

\begin{figure*}
    \centering
    \includegraphics[width=0.46\textwidth]{ 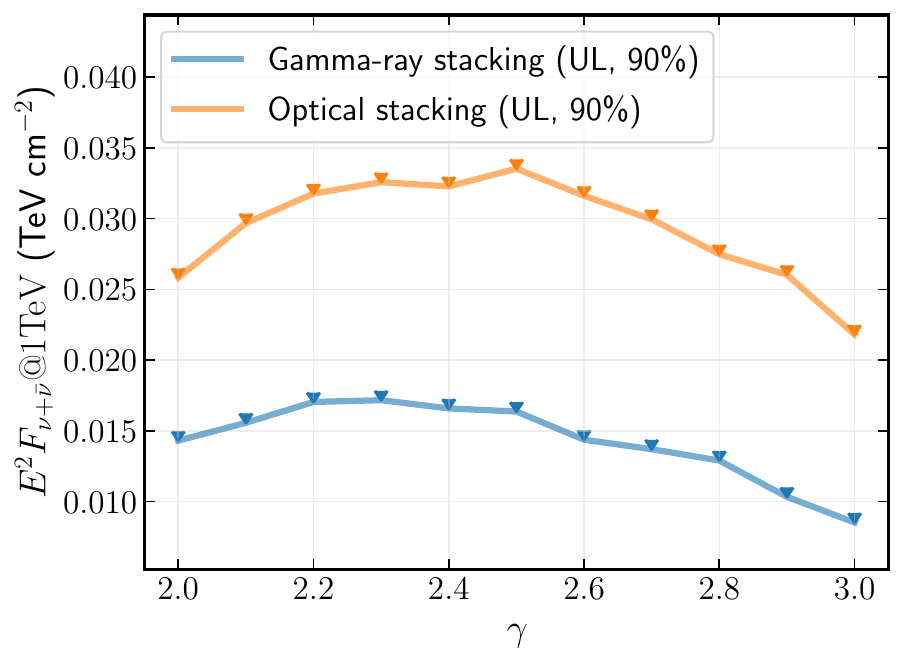}
    \includegraphics[width=0.44\textwidth]{ 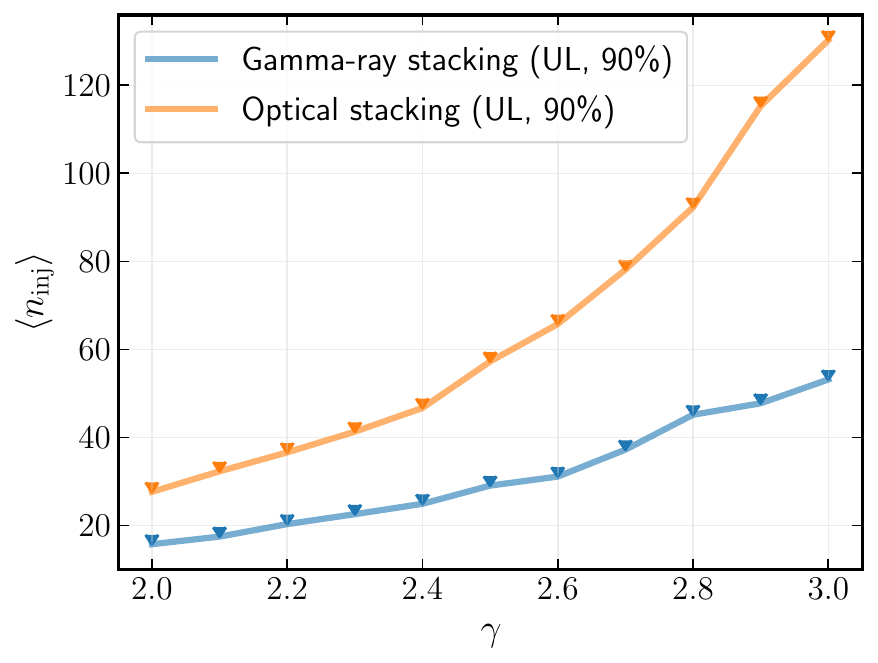}
    \caption{Upper limits on the stacked neutrino fluxes for the gamma-ray stacking (blue) and optical stacking (orange) weighting schemes. The left plot shows the energy-scaled, time-integrated, fluxes, whereas the right plot shows the corresponding expected number of signal neutrino events. Both plots show the total contributions from all sources for each weighting scheme.}
    \label{fig:stacking_limits}
\end{figure*}

As we did not detect any significant signal in the stacking analysis, we calculate upper limits for a variety of power-law spectra, for both of the weighting schemes. These limits are shown in Figure~\ref{fig:stacking_limits}, both in terms of the stacked energy-scaled time-integrated fluxes as well as in terms of the average number of signal events to which each of these fluxes correspond. These limits are at the level of the sensitivities we described in Section~\ref{sec:analysis}, which is about 1 (0.4) events from each nova, on average for the gamma-ray (optical) weighting scheme for an assumed $F_{\nu+\bar{\nu}}(E | \gamma =2.0)$ signal spectrum.

\subsection{Physical model upper limits}

In addition to calculating limits for generic unbroken power laws for the gamma-ray correlation analysis, we also inject the spectral shapes from the models developed in~\cite{Bednarek:2022vey}, which suggests that there could be neutrino fluxes at a level larger than would be naively assumed from the levels of the gamma-ray fluxes. Even though the likelihood assumes a power law in the signal PDF, we can calculate an upper limit on these model fluxes by keeping the likelihood fixed but injecting the model fluxes with various normalizations until the resulting expected test-statistic distributions are distinguishable from our observed result at 90\% confidence. We show these limits on the model fluxes in addition to the power-law upper limits in Figure~\ref{fig:ul_panel}.

For the cases of the fluxes from \cite{Bednarek:2022vey}, the spectral shapes are a function of the maximum proton energy as well as the magnetic field strength on the surface of the white dwarf at the equator. We chose the physical parameters corresponding to the tested model in \cite{Bednarek:2022vey}, which yields the highest number of events in the GRECO sample. We show two different maximum proton energies, $E_{\mathrm{max}}$, for the limits shown in Figure~\ref{fig:ul_panel} and used a nominal $B$-field value of $10^{8}$~G, although the interpretation does not change much when varying the parameters of the model and comparing to our analysis results. In total, depending on the exact physical parameters of the nova, the models of \cite{Bednarek:2022vey} predict that the GRECO Astronomy data sample could contain $\mathcal{O}(0.01)$ neutrinos from the weakest novae and $\mathcal{O}(1)$ neutrinos from the brightest novae, in the most optimistic scenarios. Our limits, on the other hand, correspond to higher fluxes that correspond to a larger number of signal events. These limits are functions of the source position, time window, as well as the observed test statistic for each nova. For novae with shorter gamma-ray detections or with smaller observed test statistics, such as V1707~Sco, V392~Per, V339~Del, V745~Sco, or V407~Lup, we find that these limits correspond to an expectation of 10\textendash 30 signal events, whereas for the other novae in the sample, the fluxes must be large enough to give an expectation of 50\textendash 200 signal events. As these upper limits are above the modeled fluxes from each nova, we cannot constrain these models. 

\section{Discussion and Conclusion}
\label{sec:discussion}
We report on the first search for neutrino emission from novae in the $\sim$ GeV to $\sim 10$~TeV energy range. Although there is promising evidence that novae are hadronic particle accelerators, the level of any corresponding neutrino flux must be at a level less than that which we are sensitive to with IceCube-DeepCore and current analysis techniques. Current models predict that, with current neutrino detectors, the novae that have been detected with photons in the last decade could be emitting neutrinos at levels that would result in $\mathcal{O}(0.01-1)$ detectable neutrinos  in IceCube-DeepCore per nova. This is below our sensitivity, even when stacking together contributions from multiple novae.

In Section~\ref{sec:novae_sample}, we discussed that we could only incorporate novae from 2012\textendash 2020 into our sample, because newer low-energy data from IceCube-DeepCore were not yet available. As a result of this, we were not able to include nova RS Ophiuchi, the first nova detected by ground-based gamma-ray observatories, into this analysis. It is worth noting that because there was very high-energy gamma-ray emission detected from this object, a separate analysis~\citep{2021ATel14851....1P} was performed to search for higher-energy neutrino emission from nova RS Ophiuchi by analyzing data available from IceCube with low latency~\citep{IceCube:2020mzw}.That search found no evidence for neutrino emission from nova RS Ophiuchi. However, that search used a data sample in which 90\% of detected events from a source at the location of nova RS Ophiuchi with an unbroken $F_{\nu+\bar{\nu}} \propto E^{-2}$ power-law spectrum would have true neutrino energies between 2~TeV and 10~PeV. This exceeds the energy range expected for neutrino emission given the detected electromagnetic emission~\citep[their Figure 12]{MAGIC:2022rmr}. 
Using the neutrino flux predictions from~\cite{MAGIC:2022rmr}, the analysis presented in this work would still not be sensitive to such a flux. Future analyses that could incorporate nova RS Ophiuchi could constrain scenarios in which the neutrino flux exceeds the observed gamma-ray flux. 

Future improvements to $\mathcal{O}(\mathrm{GeV})$ neutrino detectors could drastically change the prospects for detecting neutrinos from novae. One of the chief challenges with neutrino detection from novae is that many of the predicted spectra are expected to peak at the lowest energies to which neutrino detectors are sensitive. However, planned detectors such as the IceCube-Upgrade~\citep{Ishihara:2019aao} or the ORCA configuration of KM3NeT~\citep{KM3NeT:2021dsa} will represent a significant improvement in effective area between 1-10~GeV with respect to the GRECO Astronomy sample used here. The IceCube-Upgrade~\citep{Ishihara:2019aao} will also provide improved angular resolution for >10 GeV events. Additionally, searching for neutrino emission on longer timescales, especially when stacking, becomes difficult with low-energy events. There are already improvements being made to low-energy reconstructions with IceCube-DeepCore~\citep{IceCube:2022kff}, and reconstructions with next-generation observatories have the promise of being even better due to an enhanced understanding of systematic uncertainties. Combining these improvements -- enhanced effective area at the lowest energies to boost signal and better background discrimination from better event reconstructions which allow for longer integration times -- have the promise of greatly enhancing sensitivity to low-energy transients such as novae.

The ability to look at longer timescales is especially important in light of the recent H.E.S.S. observations of nova RS Ophiuchi~\citep{HESS:2022qap}. These observations suggested that particles that are accelerated to the highest energies only attain these energies a couple of days after the peak observed in GeV gamma rays or in optical bands. While the individual catalog analysis searched for emission on timescales of multiple days for most novae, the stacked analysis was limited to a time window 1 day in duration because of the overwhelming atmospheric backgrounds. Future analyses that feature better spatially reconstructed events could reduce the effective background and have a boosted sensitivity when searching for neutrino emission correlated with this TeV peak in the gamma-ray spectra.

Even with these improvements, it may be the case that we will need a particularly nearby or bright nova to conclusively discriminate between leptonic or hadronic models of particle acceleration in novae, or to potentially discriminate between neutrino emission models that suggest neutrino emission internal to the shocks~\citep{Bednarek:2022vey} or that would predict similar levels of neutrino and gamma-ray emission~\citep{Fang:2020bkm}. There are several nearby recurrent novae, eg. T Coronae Borealis, which has an estimated distance that is approximately 3 times nearer than the best-fit distance of $\sim 2.4$~kpc for nova RS Ophiuchi. If a future nova outburst from this system were to occur and have an emitted energy similar to RS Ophiuichi, the potential neutrino flux from this object could be detectable, especially with next generation detectors.

\section{Acknowledgements}
The IceCube collaboration acknowledges the significant contributions to this manuscript from Alex Pizzuto, Jessie Thwaites, and Justin Vandenbroucke. The authors gratefully acknowledge the support from the following agencies and institutions: USA – U.S. National Science Foundation-Office of Polar Programs, U.S. National Science Foundation-Physics Division, U.S. National Science Foundation-EPSCoR, Wisconsin Alumni Research Foundation, Center for High Throughput Computing (CHTC) at the University of Wisconsin–Madison, Open Science Grid (OSG), Advanced Cyberinfrastructure Coordination Ecosystem: Services \& Support (ACCESS), Frontera computing project at the Texas Advanced Computing Center, U.S. Department of Energy-National Energy Research Scientific Computing Center, Particle astrophysics research computing center at the University of Maryland, Institute for Cyber-Enabled Research at Michigan State University, and Astroparticle physics computational facility at Marquette University; Belgium – Funds for Scientific Research (FRS-FNRS and FWO), FWO Odysseus and Big Science programmes, and Belgian Federal Science Policy Office (Belspo); Germany – Bundesministerium für Bildung und Forschung (BMBF), Deutsche Forschungsgemeinschaft (DFG), Helmholtz Alliance for Astroparticle Physics (HAP), Initiative and Networking Fund of the Helmholtz Association, Deutsches Elektronen Synchrotron (DESY), and High Performance Computing cluster of the RWTH Aachen; Sweden – Swedish Research Council, Swedish Polar Research Secretariat, Swedish National Infrastructure for Computing (SNIC), and Knut and Alice Wallenberg Foundation; European Union – EGI Advanced Computing for research; Australia – Australian Research Council; Canada – Natural Sciences and Engineering Research Council of Canada, Calcul Québec, Compute Ontario, Canada Foundation for Innovation, WestGrid, and Compute Canada; Denmark – Villum Fonden, Carlsberg Foundation, and European Commission; New Zealand – Marsden Fund; Japan – Japan Society for Promotion of Science (JSPS) and Institute for Global Prominent Research (IGPR) of Chiba University; Korea – National Research Foundation of Korea (NRF); Switzerland – Swiss National Science Foundation (SNSF); United Kingdom – Department of Physics, University of Oxford. 

\bibliography{references}{}
\bibliographystyle{aasjournal}

\appendix
\section{Nova catalog}
\label{sec:catalog_appendix}
\input{nova_catalog_table}
\onecolumngrid
\clearpage
\section{Event Selection}
\label{app:greco}
\input{GRECO-supp}
\end{document}

%% file: authorlist120122.tex
\affiliation{III. Physikalisches Institut, RWTH Aachen University, D-52056 Aachen, Germany}
\affiliation{Department of Physics, University of Adelaide, Adelaide, 5005, Australia}
\affiliation{Dept. of Physics and Astronomy, University of Alaska Anchorage, 3211 Providence Dr., Anchorage, AK 99508, USA}
\affiliation{Dept. of Physics, University of Texas at Arlington, 502 Yates St., Science Hall Rm 108, Box 19059, Arlington, TX 76019, USA}
\affiliation{CTSPS, Clark-Atlanta University, Atlanta, GA 30314, USA}
\affiliation{School of Physics and Center for Relativistic Astrophysics, Georgia Institute of Technology, Atlanta, GA 30332, USA}
\affiliation{Dept. of Physics, Southern University, Baton Rouge, LA 70813, USA}
\affiliation{Dept. of Physics, University of California, Berkeley, CA 94720, USA}
\affiliation{Lawrence Berkeley National Laboratory, Berkeley, CA 94720, USA}
\affiliation{Institut f{\"u}r Physik, Humboldt-Universit{\"a}t zu Berlin, D-12489 Berlin, Germany}
\affiliation{Fakult{\"a}t f{\"u}r Physik {\&} Astronomie, Ruhr-Universit{\"a}t Bochum, D-44780 Bochum, Germany}
\affiliation{Universit{\'e} Libre de Bruxelles, Science Faculty CP230, B-1050 Brussels, Belgium}
\affiliation{Vrije Universiteit Brussel (VUB), Dienst ELEM, B-1050 Brussels, Belgium}
\affiliation{Department of Physics and Laboratory for Particle Physics and Cosmology, Harvard University, Cambridge, MA 02138, USA}
\affiliation{Dept. of Physics, Massachusetts Institute of Technology, Cambridge, MA 02139, USA}
\affiliation{Dept. of Physics and The International Center for Hadron Astrophysics, Chiba University, Chiba 263-8522, Japan}
\affiliation{Department of Physics, Loyola University Chicago, Chicago, IL 60660, USA}
\affiliation{Dept. of Physics and Astronomy, University of Canterbury, Private Bag 4800, Christchurch, New Zealand}
\affiliation{Dept. of Physics, University of Maryland, College Park, MD 20742, USA}
\affiliation{Dept. of Astronomy, Ohio State University, Columbus, OH 43210, USA}
\affiliation{Dept. of Physics and Center for Cosmology and Astro-Particle Physics, Ohio State University, Columbus, OH 43210, USA}
\affiliation{Niels Bohr Institute, University of Copenhagen, DK-2100 Copenhagen, Denmark}
\affiliation{Dept. of Physics, TU Dortmund University, D-44221 Dortmund, Germany}
\affiliation{Dept. of Physics and Astronomy, Michigan State University, East Lansing, MI 48824, USA}
\affiliation{Dept. of Physics, University of Alberta, Edmonton, Alberta, Canada T6G 2E1}
\affiliation{Erlangen Centre for Astroparticle Physics, Friedrich-Alexander-Universit{\"a}t Erlangen-N{\"u}rnberg, D-91058 Erlangen, Germany}
\affiliation{Physik-department, Technische Universit{\"a}t M{\"u}nchen, D-85748 Garching, Germany}
\affiliation{D{\'e}partement de physique nucl{\'e}aire et corpusculaire, Universit{\'e} de Gen{\`e}ve, CH-1211 Gen{\`e}ve, Switzerland}
\affiliation{Dept. of Physics and Astronomy, University of Gent, B-9000 Gent, Belgium}
\affiliation{Dept. of Physics and Astronomy, University of California, Irvine, CA 92697, USA}
\affiliation{Karlsruhe Institute of Technology, Institute for Astroparticle Physics, D-76021 Karlsruhe, Germany }
\affiliation{Karlsruhe Institute of Technology, Institute of Experimental Particle Physics, D-76021 Karlsruhe, Germany }
\affiliation{Dept. of Physics, Engineering Physics, and Astronomy, Queen's University, Kingston, ON K7L 3N6, Canada}
\affiliation{Department of Physics {\&} Astronomy, University of Nevada, Las Vegas, NV, 89154, USA}
\affiliation{Nevada Center for Astrophysics, University of Nevada, Las Vegas, NV 89154, USA}
\affiliation{Dept. of Physics and Astronomy, University of Kansas, Lawrence, KS 66045, USA}
\affiliation{Department of Physics and Astronomy, UCLA, Los Angeles, CA 90095, USA}
\affiliation{Centre for Cosmology, Particle Physics and Phenomenology - CP3, Universit{\'e} catholique de Louvain, Louvain-la-Neuve, Belgium}
\affiliation{Department of Physics, Mercer University, Macon, GA 31207-0001, USA}
\affiliation{Dept. of Astronomy, University of Wisconsin{\textendash}Madison, Madison, WI 53706, USA}
\affiliation{Dept. of Physics and Wisconsin IceCube Particle Astrophysics Center, University of Wisconsin{\textendash}Madison, Madison, WI 53706, USA}
\affiliation{Institute of Physics, University of Mainz, Staudinger Weg 7, D-55099 Mainz, Germany}
\affiliation{Department of Physics, Marquette University, Milwaukee, WI, 53201, USA}
\affiliation{Institut f{\"u}r Kernphysik, Westf{\"a}lische Wilhelms-Universit{\"a}t M{\"u}nster, D-48149 M{\"u}nster, Germany}
\affiliation{Bartol Research Institute and Dept. of Physics and Astronomy, University of Delaware, Newark, DE 19716, USA}
\affiliation{Dept. of Physics, Yale University, New Haven, CT 06520, USA}
\affiliation{Columbia Astrophysics and Nevis Laboratories, Columbia University, New York, NY 10027, USA}
\affiliation{Dept. of Physics, University of Oxford, Parks Road, Oxford OX1 3PU, UK}
\affiliation{Dipartimento di Fisica e Astronomia Galileo Galilei, Universit{\`a} Degli Studi di Padova, 35122 Padova PD, Italy}
\affiliation{Dept. of Physics, Drexel University, 3141 Chestnut Street, Philadelphia, PA 19104, USA}
\affiliation{Physics Department, South Dakota School of Mines and Technology, Rapid City, SD 57701, USA}
\affiliation{Dept. of Physics, University of Wisconsin, River Falls, WI 54022, USA}
\affiliation{Dept. of Physics and Astronomy, University of Rochester, Rochester, NY 14627, USA}
\affiliation{Department of Physics and Astronomy, University of Utah, Salt Lake City, UT 84112, USA}
\affiliation{Oskar Klein Centre and Dept. of Physics, Stockholm University, SE-10691 Stockholm, Sweden}
\affiliation{Dept. of Physics and Astronomy, Stony Brook University, Stony Brook, NY 11794-3800, USA}
\affiliation{Dept. of Physics, Sungkyunkwan University, Suwon 16419, Korea}
\affiliation{Institute of Physics, Academia Sinica, Taipei, 11529, Taiwan}
\affiliation{Dept. of Physics and Astronomy, University of Alabama, Tuscaloosa, AL 35487, USA}
\affiliation{Dept. of Astronomy and Astrophysics, Pennsylvania State University, University Park, PA 16802, USA}
\affiliation{Dept. of Physics, Pennsylvania State University, University Park, PA 16802, USA}
\affiliation{Dept. of Physics and Astronomy, Uppsala University, Box 516, S-75120 Uppsala, Sweden}
\affiliation{Dept. of Physics, University of Wuppertal, D-42119 Wuppertal, Germany}
\affiliation{Deutsches Elektronen-Synchrotron DESY, Platanenallee 6, 15738 Zeuthen, Germany }

\author[0000-0001-6141-4205]{R. Abbasi}
\affiliation{Department of Physics, Loyola University Chicago, Chicago, IL 60660, USA}

\author[0000-0001-8952-588X]{M. Ackermann}
\affiliation{Deutsches Elektronen-Synchrotron DESY, Platanenallee 6, 15738 Zeuthen, Germany }

\author{J. Adams}
\affiliation{Dept. of Physics and Astronomy, University of Canterbury, Private Bag 4800, Christchurch, New Zealand}

\author{N. Aggarwal}
\affiliation{Dept. of Physics, University of Alberta, Edmonton, Alberta, Canada T6G 2E1}

\author[0000-0003-2252-9514]{J. A. Aguilar}
\affiliation{Universit{\'e} Libre de Bruxelles, Science Faculty CP230, B-1050 Brussels, Belgium}

\author[0000-0003-0709-5631]{M. Ahlers}
\affiliation{Niels Bohr Institute, University of Copenhagen, DK-2100 Copenhagen, Denmark}

\author[0000-0002-9534-9189]{J.M. Alameddine}
\affiliation{Dept. of Physics, TU Dortmund University, D-44221 Dortmund, Germany}

\author{A. A. Alves Jr.}
\affiliation{Karlsruhe Institute of Technology, Institute for Astroparticle Physics, D-76021 Karlsruhe, Germany }

\author{N. M. Amin}
\affiliation{Bartol Research Institute and Dept. of Physics and Astronomy, University of Delaware, Newark, DE 19716, USA}

\author{K. Andeen}
\affiliation{Department of Physics, Marquette University, Milwaukee, WI, 53201, USA}

% ...

\date{\today}

\collaboration{388}{IceCube Collaboration}

%% file: gamma_ray_results.tex
\begin{table*}
\centering
\caption{Results from the Gamma-ray Correlation Analysis.}
\begin{tabular}{ccccccccccc}
\toprule
Name &                 $\alpha$ & $\delta$ & MJD$_{\mathrm{start}}$ & MJD$_{\mathrm{stop}}$ &  $\Delta$T & $\mathcal{TS}$ & $\hat{n}_{s}$ &$\hat{\gamma}$ &   Pre-trial & $E^2F_{\nu+\bar{\nu}}$ @1~TeV \\
&  &  & &  &(days)  &  & & &   $p$-value & (GeV cm$^{-2}$)\\
\midrule
V1324 Sco & 267.7$^{\circ}$ & -32.6$^{\circ}$ &                56093.0 &               56110.0 & 17.0 &         0.00 &                     0.0& -- & 1.000 & 52.9\\%
 V959 Mon &   99.9$^{\circ}$ & +5.9$^{\circ}$ &                56097.0 &               56119.0 & 22.0 &         1.14 &                  14.8& 2.75 & 0.197 & 19.5\\%
 V339 Del & 305.9$^{\circ}$& +20.8$^{\circ}$ &                56520.0 &               56547.0 & 27.0 &         0.00 &                     0.0& -- & 1.000 & 4.21\\%
V1369 Cen & 208.7$^{\circ}$& -59.2$^{\circ}$ &                56634.0 &               56672.0 & 38.0 &          0.01 &                   3.8& 4.00 & 0.070 & 125.9\\%
 V745 Sco & 268.8$^{\circ}$& -33.2$^{\circ}$ &                56694.0 &               56695.0 & 1.0 &         0.00 &                     0.0& -- & 1.000 & 14.9\\%
V1535 Sco & 255.9$^{\circ}$& -35.1$^{\circ}$ &                57064.0 &               57071.0 & 7.0 &         6.95 &                  59.9& 3.02 & 0.002 & 85.2\\%
V5668 Sgr & 279.2$^{\circ}$& -28.9$^{\circ}$ &                57105.0 &               57158.0 &  53.0 &        1.30 &                  62.0& 3.16 & 0.112 & 139.2\\%
 V407 Lup & 232.3$^{\circ}$& -44.8$^{\circ}$ &                57657.0 &               57660.0 & 3.0 &         0.00 &                     0.0& -- & 1.000 & 22.4\\%
V5855 Sgr & 272.6$^{\circ}$& -27.5$^{\circ}$ &                57686.0 &               57712.0 & 26.0 &           0.00 &                     0.0& -- & 1.000 & 56.6\\%
V5856 Sgr & 275.2$^{\circ}$& -28.4$^{\circ}$ &                57700.0 &               57715.0 & 15.0 &          5.13 &                  40.3& 2.81 & 0.015 & 94.0\\%
 V549 Vel & 132.6$^{\circ}$& -47.8$^{\circ}$ &                58037.0 &               58070.0 & 33.0 &          0.22 &                  22.4& 4.00 & 0.062 & 103.2\\%
 V357 Mus & 171.6$^{\circ}$& -65.5$^{\circ}$ &                58129.0 &               58156.0 & 27.0 &          0.01 &                   5.0& 4.00 & 0.115 & 104.6\\%
 V906 Car & 159.1$^{\circ}$& -59.6$^{\circ}$ &                58216.0 &               58239.0 & 23.0 &         0.00 &                     0.0& -- & 1.000 & 73.0\\%
 V392 Per &  70.8$^{\circ}$& +47.4$^{\circ}$ &                58238.0 &               58246.0 & 8.0 &          0.88 &                  15.4& 3.64 & 0.373 & 3.84\\%
V3890 Sgr & 277.7$^{\circ}$& -24.0$^{\circ}$ &                58718.0 &               58739.0 & 21.0 &          0.00 &                     0.0& -- & 1.000 & 45.1\\%
V1707 Sco & 264.3$^{\circ}$& -35.2$^{\circ}$ &                58740.0 &               58744.0 & 4.0 &          0.00 &                     0.0& -- & 1.000 & 21.2\\%
\bottomrule
\end{tabular}
\label{tab:gamma_results}

\raggedright
\vspace{2pt}
\textbf{Note.} MJD$_{\mathrm{start}}$ and MJD$_{\mathrm{stop}}$ represent the beginning and end of when a nova was detected by \textit{Fermi}-LAT, respectively. This is the same time window used for the neutrino search. The observed test statistic ($\mathcal{TS}$), best-fit number of signal events ($\hat{n}_{s}$) and best-fit spectral index ($\hat{\gamma}$) for the neutrino search from each nova are given here. The $p$-values shown are before accounting for the factor accrued from performing multiple searches. The final column gives the time-integrated flux upper limit for each nova, assuming a power-law spectrum with $\gamma=2.0$ as defined in Equation \ref{eq:dnde} and including the systematics discussed in Section \ref{sec:systematics}.
\end{table*}

%% file: nova_catalog_table.tex
\startlongtable
\begin{deluxetable*}{lllllcl}
\tablecaption{Catalog of Galactic Novae Temporally Coincident with Our Neutrino Data Sample Used in This Analysis. }
\tablehead{
\colhead{Name} & \colhead{Peak Date} & \colhead{$\alpha$} & \colhead{$\delta$} & \colhead{Peak Mag.} & \colhead{$\gamma$-detected} & \colhead{Reference}
}
\startdata
OGLE-2012-NOVA-01 & 2012-05-05 & 269.20$^{\circ}$ & -27.23$^{\circ}$ & 12.5 (I) & -- & 1 \\
V2677 Oph & 2012-05-20 & 264.99$^{\circ}$ & -24.78$^{\circ}$ & 9.5 & -- & 1 \\
V1324 Sco & 2012-06-20 & 267.72$^{\circ}$ & -32.62$^{\circ}$ & 9.8 & \checkmark & 2, 1 \\
V959 Mon & 2012-06-24 & 99.91$^{\circ}$ & +5.90$^{\circ}$ & 5.0 & \checkmark & 2, 1 \\
V5591 Sgr & 2012-06-27 & 268.10$^{\circ}$ & -21.44$^{\circ}$ & 8.9 & -- & 1 \\
OGLE-2011-BLG-1444 & 2012-07-01 & 267.58$^{\circ}$ & -33.65$^{\circ}$ & 10.2 (I) & -- & 1 \\
V5592 Sgr & 2012-07-07 & 275.11$^{\circ}$ & -27.74$^{\circ}$ & 7.7 & -- & 1 \\
V5593 Sgr & 2012-07-23 & 274.90$^{\circ}$ & -19.12$^{\circ}$ & 11.0 & -- & 1 \\
V1724 Aql & 2012-10-20 & 283.15$^{\circ}$ & -0.32$^{\circ}$ & 11.2 & -- & 1 \\
V809 Cep & 2013-02-02 & 347.02$^{\circ}$ & +60.78$^{\circ}$ & 9.8 & -- & 1 \\
OGLE-2013-NOVA-04 & 2013-04-14 & 270.34$^{\circ}$ & -25.31$^{\circ}$ & 13.0 (I) & -- & 1 \\
V1533 Sco & 2013-06-03 & 263.50$^{\circ}$ & -36.11$^{\circ}$ & 11.0 & -- & 1 \\
V339 Del & 2013-08-16 & 305.88$^{\circ}$ & +20.77$^{\circ}$ & 4.5 & \checkmark & 2, 1 \\
V1830 Aql & 2013-10-28 & 285.64$^{\circ}$ & +3.26$^{\circ}$ & 13.5 & -- & 1 \\
V556 Ser & 2013-11-24 & 272.26$^{\circ}$ & -11.21$^{\circ}$ & 11.7 & -- & 1 \\
V1369 Cen & 2013-12-14 & 208.69$^{\circ}$ & -59.15$^{\circ}$ & 3.3 & \checkmark & 3, 1 \\
V5666 Sgr & 2014-01-26 & 276.29$^{\circ}$ & -22.61$^{\circ}$ & 8.7 & -- & 1 \\
V745 Sco & 2014-02-06 & 268.84$^{\circ}$ & -33.25$^{\circ}$ & 8.7 & \checkmark & 1 \\
OGLE-2014-NOVA-01 & 2014-02-16 & 268.83$^{\circ}$ & -23.40$^{\circ}$ & 15.0 (I) & -- & 1 \\
V962 Cep & 2014-03-14 & 313.59$^{\circ}$ & +60.28$^{\circ}$ & 11.0 & -- & 1 \\
V1534 Sco & 2014-03-26 & 258.94$^{\circ}$ & -31.48$^{\circ}$ & 10.1 & -- & 1 \\
V2659 Cyg & 2014-04-10 & 305.43$^{\circ}$ & +31.06$^{\circ}$ & 9.4 & -- & 1 \\
V1535 Sco & 2015-02-11 & 255.86$^{\circ}$ & -35.07$^{\circ}$ & 8.2 & \checkmark & 1 \\
V5667 Sgr & 2015-02-25 & 273.60$^{\circ}$ & -25.91$^{\circ}$ & 9.0 & -- & 1 \\
V5668 Sgr & 2015-03-21 & 279.24$^{\circ}$ & -28.93$^{\circ}$ & 4.3 & \checkmark & 3, 1 \\
V1658 Sco & 2015-04-03 & 267.05$^{\circ}$ & -32.59$^{\circ}$ &  12.7 & -- & 4 \\
V2944 Oph & 2015-04-14 & 262.30$^{\circ}$ & -18.77$^{\circ}$ & 9.0 & -- & 1 \\
V5669 Sgr & 2015-09-27 & 270.89$^{\circ}$ & -28.27$^{\circ}$ & 8.7 & -- & 1 \\
V2949 Oph & 2015-10-12 & 263.70$^{\circ}$ & -24.15$^{\circ}$ & 11.7 & -- & 1 \\
V1831 Aql & 2015-10-13 & 290.46$^{\circ}$ & +15.15$^{\circ}$ & 13.8 & -- & 1 \\
V5850 Sgr & 2015-10-31 & 275.75$^{\circ}$ & -19.24$^{\circ}$ & 11.3 & -- & 1 \\
V3661 Oph & 2016-03-13 & 263.96$^{\circ}$ & -29.57$^{\circ}$ &  10.6 & -- & 5 \\
V555 Nor & 2016-04-02 & 235.44$^{\circ}$ & -53.14$^{\circ}$ &  12.43 & -- & \atel{10740} \\
V1655 Sco & 2016-06-12 & 264.58$^{\circ}$ & -37.42$^{\circ}$ &  10.4 & -- & 6 \\
V5853 Sgr & 2016-08-12 & 270.28$^{\circ}$ & -26.53$^{\circ}$ &  10.7 & -- & 5 \\
V1659 Sco & 2016-09-07 & 265.74$^{\circ}$ & -33.43$^{\circ}$ &  12.3 & -- & 7 \\
V611 Sct & 2016-09-07 & 276.37$^{\circ}$ & -9.79$^{\circ}$ &  13.4 & -- & 7 \\
V1656 Sco & 2016-09-11 & 260.71$^{\circ}$ & -31.98$^{\circ}$ &  11.4 & -- & 5 \\
V407 Lup & 2016-09-26 & 232.26$^{\circ}$ & -44.83$^{\circ}$ &  5.6 & \checkmark & \atel{9594}, 8, 5 \\
V5855 Sgr & 2016-10-24 & 272.62$^{\circ}$ & -27.50$^{\circ}$ &  9.8 & \checkmark & 9 \\
V5856 Sgr & 2016-11-08 & 275.22$^{\circ}$ & -28.37$^{\circ}$ &  5.4 & \checkmark & 10 \\
V3662 Oph & 2017-04-30 & 264.94$^{\circ}$ & -24.97$^{\circ}$ &  14.1 & -- & \atel{10367} \\
V1405 Cen & 2017-05-17 & 200.23$^{\circ}$ & -63.71$^{\circ}$ &  10.9 & -- & \atel{10387} \\
V612 Sct & 2017-06-20 & 277.94$^{\circ}$ & -14.32$^{\circ}$ &  8.5 & -- & 7 \\
V549 Vel & 2017-10-09 & 132.62$^{\circ}$ & -47.76$^{\circ}$ &  9.3 & \checkmark & \atel{109677}, 7, 13 \\
V1660 Sco & 2017-10-14 & 262.64$^{\circ}$ & -31.10$^{\circ}$ &  13.0 & -- & \atel{10850} \\
V1661 Sco & 2018-01-17 & 259.53$^{\circ}$ & -32.07$^{\circ}$ &  10.7 & -- & 5 \\
V357 Mus & 2018-01-17 & 171.55$^{\circ}$ & -65.52$^{\circ}$ &  6.7 & \checkmark & \atel{11201}, 8 \\
V1662 Sco & 2018-02-11 & 252.21$^{\circ}$ & -44.95$^{\circ}$ &  9.8 & -- & 6 \\
V3664 Oph & 2018-02-13 & 261.17$^{\circ}$ & -24.36$^{\circ}$ &  12.8 & -- & 6 \\
V1663 Sco & 2018-03-01 & 255.95$^{\circ}$ & -38.28$^{\circ}$ &  11.4 & -- & 6 \\
V3665 Oph & 2018-03-11 & 258.51$^{\circ}$ & -28.82$^{\circ}$ &  9.3 & -- & 6 \\
V435 CMa & 2018-04-07 & 108.44$^{\circ}$ & -21.21$^{\circ}$ &  12.1 & -- & 6 \\
V5857 Sgr & 2018-04-10 & 271.04$^{\circ}$ & -18.07$^{\circ}$ &  10.8 & -- & 6 \\
V906 Car & 2018-04-14 & 159.06$^{\circ}$ & -59.60$^{\circ}$ &  6.62 & \checkmark & 8, 12 \\
V392 Per & 2018-04-29 & 70.84$^{\circ}$ & +47.36$^{\circ}$ &  6.2 & \checkmark & \atel{11590}, 11, 6 \\
V408 Lup & 2018-06-09 & 234.68$^{\circ}$ & -47.74$^{\circ}$ &  8.98 & -- & 5 \\
V613 Sct & 2018-07-01 & 277.35$^{\circ}$ & -14.51$^{\circ}$ &  10.4 & -- & 6 \\
V3666 Oph & 2018-08-11 & 265.60$^{\circ}$ & -20.89$^{\circ}$ &  9.0 & -- & 6 \\
V556 Nor & 2018-10-14 & 243.64$^{\circ}$ & -53.50$^{\circ}$ &  11.0 & -- & 6 \\
V1706 Sco & 2019-05-14 & 256.89$^{\circ}$ & -36.14$^{\circ}$ &  13.0 & -- & 7 \\
V2860 Ori & 2019-08-08 & 92.49$^{\circ}$ & +12.21$^{\circ}$ &  9.4 & -- & 6 \\
V569 Vul & 2019-08-18 & 298.03$^{\circ}$ & +27.71$^{\circ}$ &  16.3 & -- & \atel{13068} \\
V3890 Sgr & 2019-08-28 & 277.68$^{\circ}$ & -24.02$^{\circ}$ &  6.7 & \checkmark & 13, \atel{13047} \\
V1707 Sco & 2019-09-15 & 264.29$^{\circ}$ & -35.17$^{\circ}$ &  10.3 & \checkmark & 6 \\
V659 Sct & 2019-10-30 & 280.00$^{\circ}$ & -10.43$^{\circ}$ &  8.3 & -- & 6 \\
V6566 Sgr & 2020-01-30 & 269.06$^{\circ}$ & -20.72$^{\circ}$ &  11.03 & -- & 6 \\
V670 Ser & 2020-02-22 & 272.68$^{\circ}$ & -15.57$^{\circ}$ &  11.8 & -- & 6 \\
\enddata

\raggedright
\vspace{2pt}
\textbf{Note.} References indicate where optical or gamma-ray lightcurves were obtained. References: 1=\cite{Franckowiak:2017iwj}, 2=\cite{Ackermann:2014vfa}, 3=\cite{Cheung:2016cqs}, 4=\cite{2016MNRAS.461.1529A}, 5=SMARTS, 6=AAVSO, 7=\cite{Shappee:2013mna}, 8=\cite{Gordon:2020fqv}, 9=\cite{Nelson:2018hjb}, 10=\cite{Li:2017crr}, 11=\cite{Li:2020njc}, 12=\cite{Aydi:2020znu}, 13=\cite{Page:2020cea} \label{tab:novae_cat}.
\end{deluxetable*}

%% file: GRECO-supp.tex
For this work, we use a modified version of the GRECO (GeV Reconstructed Events with Containment for Oscillation) event selection, which was initially developed for the tau neutrino appearance analysis described in~\cite{IceCube:2019dqi}. The original GRECO dataset, which was referred as Analysis $\mathcal{A}$ in that reference, consisted of 3 years of data from 2012 April through 2015 May covering the full sky and all neutrino flavors. This original dataset was used in an initial untriggered all-sky search for sub-TeV transient neutrino emission \citep{IceCube:2020qls}. 

\begin{figure}[h!]
    \centering
    \includegraphics[width=0.5\textwidth]{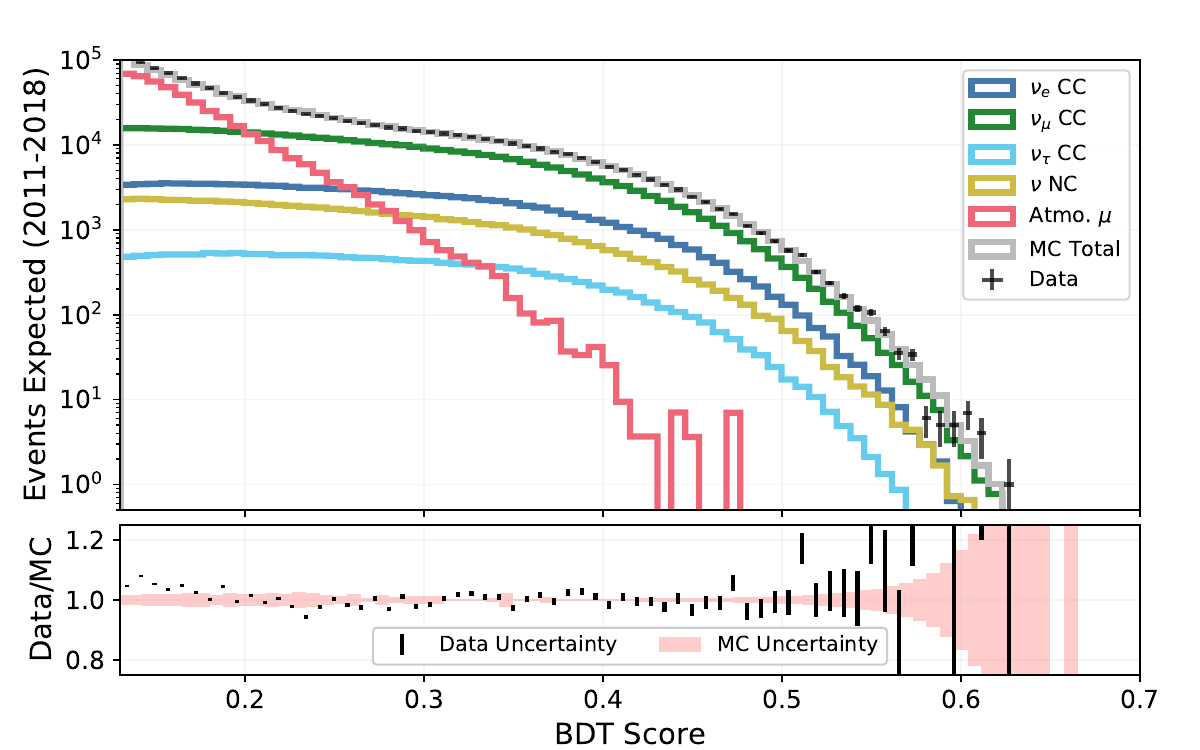}
    \caption{The expected contributions from each atmospheric background component, including neutrinos and simulations of atmospheric muons. Neutrino charged-current (CC) interactions for each neutrino flavor, neutral-current (NC) interactions for neutrinos of all flavors, and atmospheric muons are labeled in the legend, with the total of all contributions shown in gray. Error bars shown are statistical and do not include any relevant systematic uncertainties. Events with a BDT score below 0.13 are removed, yielding a sample with 60\% atmospheric neutrinos and 40\% atmospheric muons.}
    \label{fig:bdt}
\end{figure}

An updated version of the GRECO dataset is used here, incorporating updates and reoptimizations developed since the original dataset was produced to identify neutrino candidate events below about 1 TeV starting in the deep, clear ice of the DeepCore detector \citep{IceCube:2019dqi}.
Simulations of the updated dataset -- here referred to as the GRECO Astronomy dataset -- use an updated version of the GENIE neutrino generator \citep{Andreopoulos:2009rq} and include new charge calibrations for both data and simulation, an updated model of the glacial ice, and newly available simulations of atmospheric muons using an implementation of MuPAGE \citep{MUPAGE:2008} referred to as ``MuonGun.''
Thirteen variables based on the original GRECO dataset are combined into a newly trained boosted decision tree (BDT). These include charge and time-evolution variables, position of the DOM with the earliest local coincidence hit in the detector, the amount of causally connected charge deposition found in the larger IceCube detector, and preliminary directional reconstruction.
The results of the BDT training, shown in Figure~\ref{fig:bdt}, show that the data is well modeled with simulation weighted by atmospheric fluxes.

\begin{figure*}[b]
    \centering
    \includegraphics[width=0.6\textwidth]{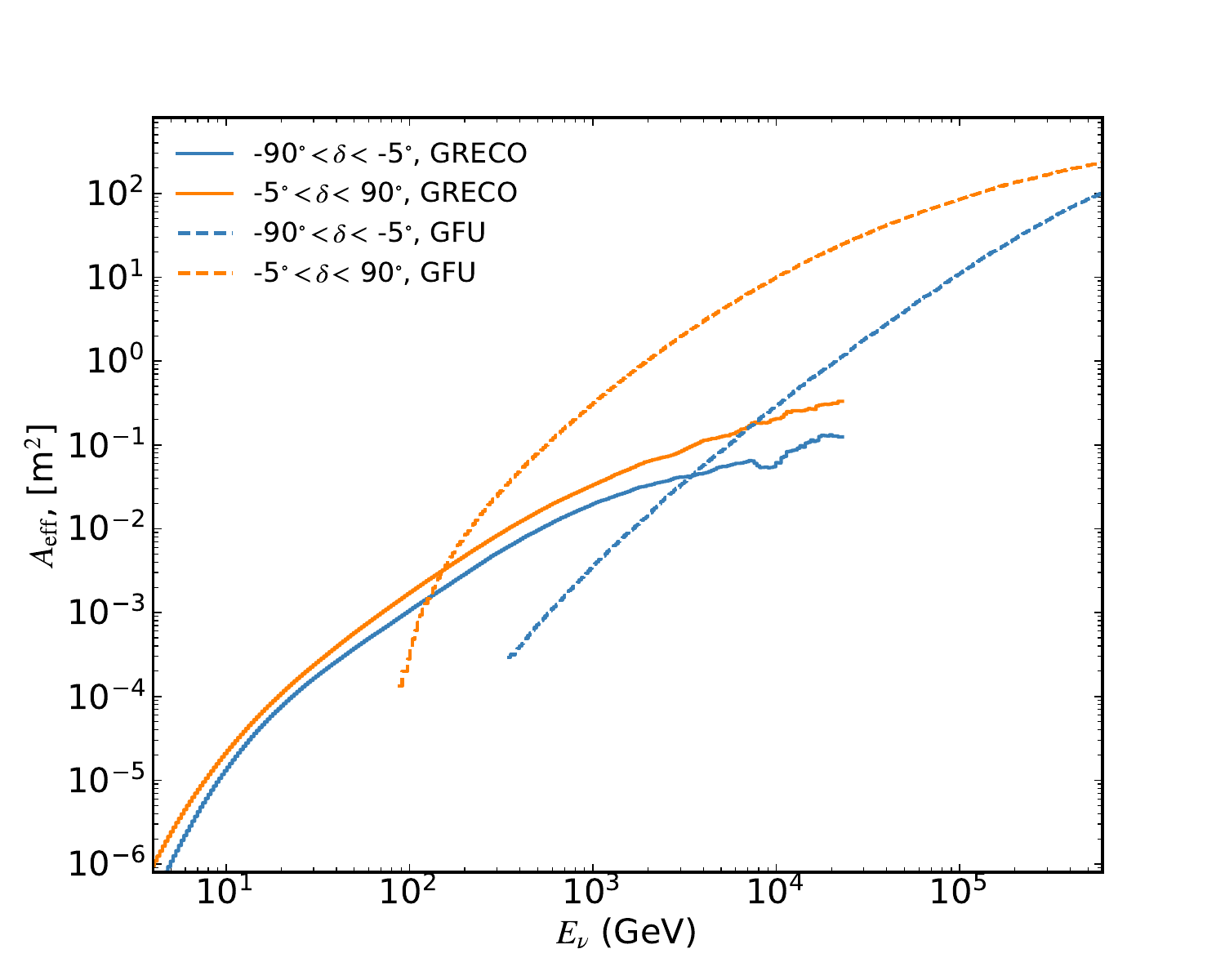}
    \caption{All-flavor, $\nu+\bar{\nu}$ averaged effective area for the low-energy GRECO event selection used for this analysis (solid lines). We include, for comparison, a higher-energy $\nu_{\mu}+\bar{\nu_\mu}$-only event selection that is often used to search for astrophysical neutrino transients (dashed lines).}
    \label{fig:eff_a}
\end{figure*}

\begin{figure*}
    \centering
    \includegraphics[width=0.92\textwidth]{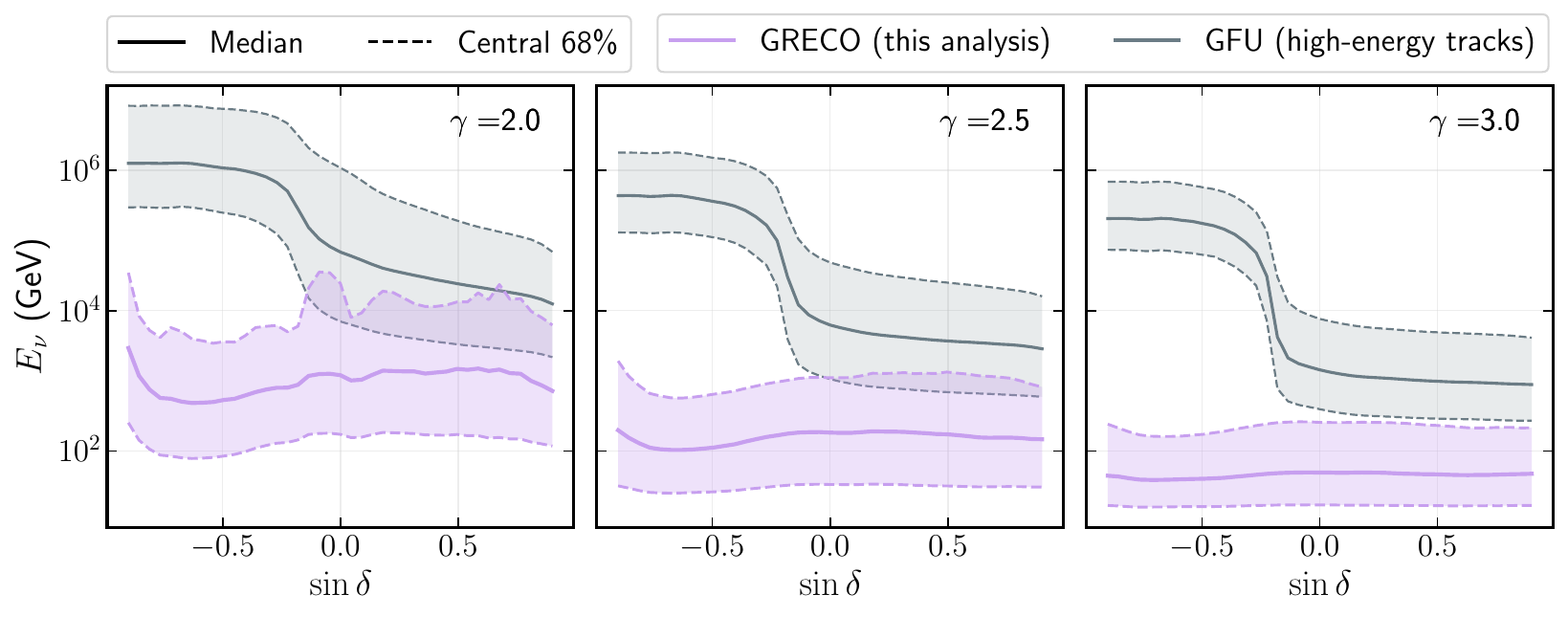}
    \caption{Central 68\% of the expected true neutrino energies for sources with $F_{\nu+\bar{\nu}} \propto E^{-2}$ (left), $E^{-2.5}$ (middle), or $E^{-3}$ (right) spectra. We compare this analysis (pink) to a higher-energy event selection (gray). This distribution only shows the distribution of true energies that are present in the dataset, although higher-energy and better reconstructed events tend to have a larger effect on the analysis sensitivity than lower-energy events.}
    \label{fig:central_energies}
\end{figure*}

The GRECO Astronomy dataset provides an improved effective area relative to the dataset used in the tau neutrino appearance search, with the largest improvements above 100 GeV. While the background rates are also higher, searches for short transient phenomena like those performed in this work are not significantly affected by the additional background contamination. 
The effective area of this sample is shown in Figure~\ref{fig:eff_a}, and we compare it against the IceCube ``Gamma-ray Follow-up'' (GFU) sample, an event selection that is often used when searching for high-energy astrophysical transients~\citep{IceCube:2016cqr}. To highlight the complementary nature of the different event selections, GRECO and GFU, we show the energies one could expect to detect from a source with various spectral hypotheses in Figure~\ref{fig:central_energies}. Although using lower-energy events to search for astrophysical neutrinos is not without limitations, as described below, it opens up a new energy range to search for multimessenger sources. 


The final sample has an average all-sky rate of 4.6~mHz, consisting of about 60\% neutrino events from cosmic-ray interactions in the atmosphere and 40\% muons from atmospheric air showers. 
The sample contains large backgrounds relative to the total expected astrophysical signal, but these backgrounds are strongly suppressed in searches for short astrophysical transients.

\begin{figure}
    \centering
    \includegraphics[width=0.45\textwidth]{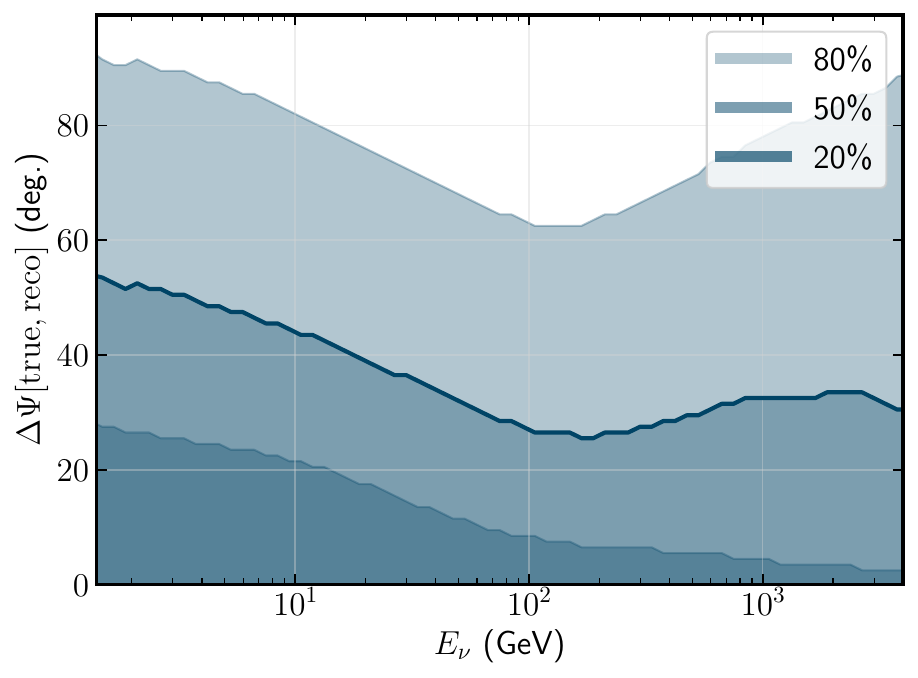}
    \caption{Distribution of the angular separation between reconstructed event direction and true neutrino direction as a function of energy, at 3 confidence levels (20\%, 50\%, 80\%). All neutrino flavors, weighted according to the per-flavor effective area, are included.}
    \label{fig:greco_err}
\end{figure}

Events observed in the detector can be classified into two distinct morphologies: tracks and cascades. 
Tracks are elongated signatures and are the result of muons traversing the detector, either from cosmic-ray interactions or from charged-current (CC) muon-neutrino interactions. 
Cascades, on the other hand, are more spherical, and are the result of neutral current interactions from all neutrino flavors or electron or tau neutrino CC interactions. 
The long lever arm of tracks allows for much better pointing resolution than for cascades.

Neutrino events starting in the detector provide a third morphology due to the physics of the interaction. 
Neutrinos interacting inside of the detector first produce a hadronic cascade due to deep inelastic scattering interactions off of the nuclei in the ice. 
Following this initial hadronic cascade, neutrinos may produce an outgoing muon track ($\nu_\mu$ charged-current), an electromagnetic shower ($\nu_e$ charged-current), a tau lepton and associated decay ($\nu_\tau$ charged-current), or a hidden outgoing neutrino (neutral current). 
The relative brightness of the initial hadronic interaction and the subsequent emission profile of the outgoing lepton degrades and, therefore, limits our ability to clearly identify the type of neutrino involved as the energy decreases. 

The GRECO Astronomy dataset contains events of relatively low energy, with low light emission, and therefore few photons reach the optical modules due to the sparseness of the array. 
A typical event in the GRECO Astronomy sample often has around 20\textendash 30 hits or less over a relatively small spatial extent. 
Any muon track contained in the GRECO Astronomy dataset also tends to be short, crossing one to two strings on average, limiting our ability to reconstruct events or differentiate low-energy tracks from cascades.

These limitations result in large directional uncertainties of the events. 
For tracks at low energies, the pointing resolution is worse than what is achieved in searches for high-energy neutrino sources. 
For comparison, in analyses that use higher-energy track events ($E_{\nu} \gtrsim 1~\mathrm{TeV}$), the median opening angle between the true neutrino direction and the reconstructed event direction is, on average, $< 1^{\circ}$. 
For lower-energy events, such as those used here, each neutrino interaction deposits much less energy in the detector, which leads to events that are much more poorly localized, with average opening angles between the true neutrino direction and the reconstructed event direction on the order of tens of degrees. 
Additionally, at lower energies, the kinematic opening angle becomes nonnegligible in comparison to the resolution of the detector. 

In Figure~\ref{fig:greco_err}, we show the average angular separation between the true neutrino direction and the reconstructed event direction for all flavors of neutrino, as both track and cascade events are included in the GRECO sample. The reconstruction algorithm was designed for the original neutrino oscillation event selection and is described at length in~\cite{IceCube:2019dqi}. The reconstruction takes a maximum likelihood approach where the underlying hypothesis assumes that every event begins with an electromagnetic or hadronic shower and then has an outgoing, finite muon originating at the same interaction vertex. Some of the other assumptions that go into this reconstruction, such as assuming that the muons are in the minimum ionizing regime, reflect the fact that the reconstruction was designed specifically for lower-energy events ($E_{\nu} \lesssim 300~\mathrm{GeV}$), and is outperformed by other reconstruction algorithms at higher energies. The mixed track and cascade hypothesis makes the likelihood an eight-dimensional parameter space, and performing the reconstruction can be extremely computationally intensive, with each event requiring about 2 minutes on average to identify a minimum.
Minimizations are performed over the full space using ``Nestle,'' a tool for posterior probabilities using a nested sampling \citep{nestle, 2009MNRAS.398.1601F, 2006ApJ...638L..51M, 2007MNRAS.378.1365S, silvia:2006, 2004AIPC..735..395S}.

Instead of using approaches that require fine scans of the likelihood space as a function of the direction~\citep{Neunhoffer:2004ha}, a more computationally efficient approach that only relies on the final event observables was used to estimate the angular uncertainty of each event. To accomplish this, a random forest was trained using a three-fold cross validation technique with neutrino simulation that contained both track-like and cascade-like events. This model outputs a singular value, $\sigma$, that describes the angular uncertainty of each event, which is used in the likelihood-based approaches implemented to search for pointlike astrophysical neutrino emission. It was found that this approach of assigning per-event angular uncertainties outperformed previous attempts to describe low-energy event angular uncertainties that grouped together all events with similar deposited energies and reconstructed zenith angles and assigned angular uncertainties based on the median opening angle between reconstructed events and their true directions from simulation~\citep{IceCube:2020qls}.

\begin{figure*}
    \centering
    \includegraphics[width=0.92\textwidth]{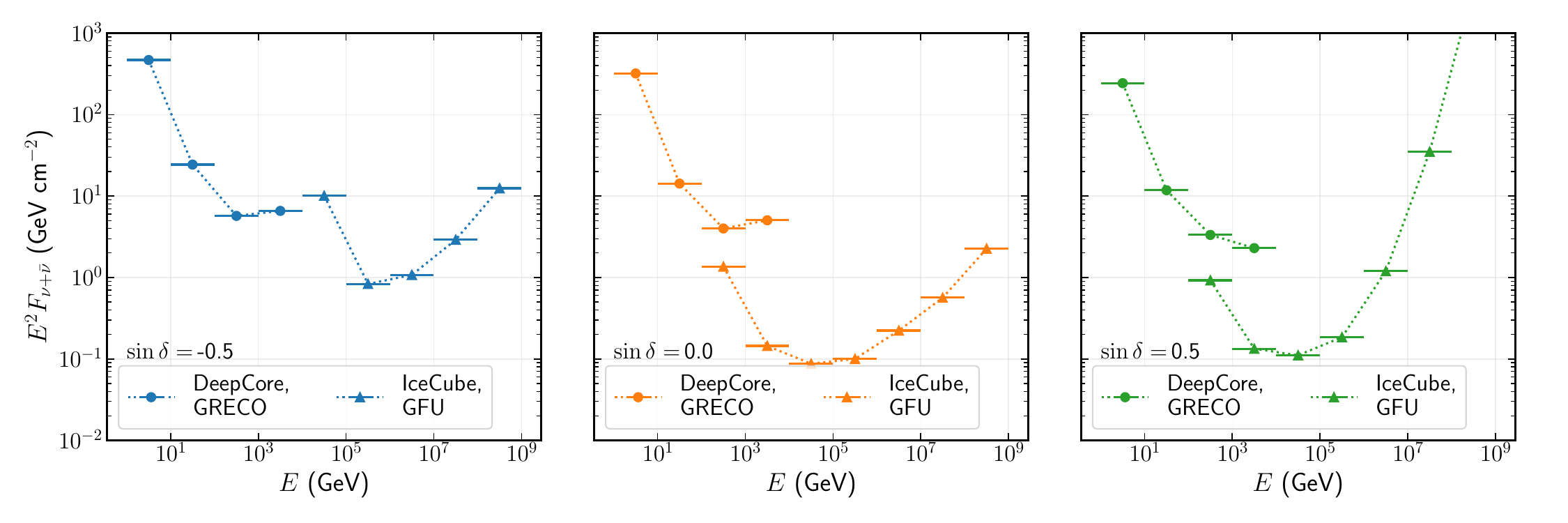}
    \caption{
    Differential sensitivities for 1000 s timescale transients using the GRECO Astronomy event selection (circles) and comparing to a higher-energy event selection (GFU, triangles). We inject a power-law spectrum with $F_{\nu+\bar{\nu}} \propto E^{-2}$ in each decadal bin. The panels show three representative declinations ($\mathrm{sin}\,\delta$) in the Southern Hemisphere, at the horizon, and in the Northern Hemisphere.}
    \label{fig:differential_sens}
\end{figure*}

In Figure \ref{fig:differential_sens}, we show differential sensitivities for this GRECO Astronomy data sample, compared to the GFU sample, consisting of higher-energy tracks. We see that the sensitivities are comparable at the energy bins where there is overlap between the two datasets. While the GFU dataset has less sensitivity at its lower-energy bins in the Southern Hemisphere, the GRECO dataset shows similar sensitivities across the entire sky. 